%% file: TPClosedEmbMining.tex
\documentclass[11pt,a4paper]{article}
\usepackage{authblk}

\usepackage{graphicx}
\usepackage{caption}
\usepackage{float}
\usepackage{algorithm}
\usepackage[noend]{algorithmic}
\graphicspath{figures/}
\usepackage{times}
\usepackage{epsfig}
\usepackage{subfigure}
\usepackage{multirow}
\usepackage{amsmath}
\usepackage{mathtools}
\usepackage{amsfonts}
\usepackage{amssymb}
\usepackage{amscd}
\newfloat{algpseudocode}{t}{lop}

\newtheorem{theorem}{Theorem}[section]

\newtheorem{definition}{Definition}[section]
\newtheorem{proposition}{Proposition}[section]
\newtheorem{example}{Example}[section]
\newtheorem{lemma}{Lemma}[section]
\newcommand{\comments}[1]{}

\begin{document}
\date{}

\title{Discovering Closed and Maximal Embedded Patterns from Large Tree Data}

\author[1]{Xiaoying Wu\thanks{xiaoying.wu@whu.edu.cn}}
\author[2]{Dimitri Theodoratos\thanks{dth@njit.edu}}
\author[3]{Nikos Mamoulis\thanks{nikos@cs.uoi.gr}}
\affil[1]{School of Computer, Wuhan University, China}
\affil[2]{New Jersey Institute of Technology, USA}
\affil[3]{University of Ioannina, Greece}

\maketitle

\begin{abstract}

Many current applications and systems produce large tree datasets and export, exchange, and represent data in tree-structured form. Extracting informative patterns from large data trees is an important research direction with multiple applications in practice. Pattern mining research initially focused on  mining induced patterns and gradually evolved into mining embedded patterns. Induced patterns cannot capture many useful relationships  hidden deeply in the datasets which can discovered by embedded patterns. A well-known problem of frequent pattern mining  is the a huge number of patterns it produces. This affects not only the efficiency but also the effectiveness of mining. A typical solution to this problem is to summarize frequent patterns through closed and maximal patterns. Previous techniques on mining closed induced tree patterns do not apply to embedded tree patterns. No previous work addresses the problem of mining closed and/or maximal embedded tree patterns, not even in the framework of mining multiple small trees.

We address the problem of summarizing embedded tree patterns extracted from large data trees. We do so by defining and mining closed and maximal embedded unordered tree patterns from a single large data tree. We design an embedded frequent pattern mining algorithm extended with a local closedness checking technique. This algorithm is called {\em closedEmbTM-prune} as it eagerly eliminates non-closed patterns. To mitigate the generation of intermediate patterns, we devise pattern search space pruning rules to proactively detect and prune branches in the pattern search space which do not correspond to closed patterns. The pruning rules are accommodated into the extended embedded pattern miner to produce a new algorithm, called {\em closedEmbTM-prune},  for mining all the closed and maximal embedded frequent patterns from large data trees. Our extensive experiments on synthetic and real large-tree datasets demonstrate that, on dense datasets, {\em closedEmbTM-prune}  not only generates a complete closed and maximal pattern set which is substantially smaller than that generated by the embedded pattern miner, but also runs much faster with negligible overhead on pattern pruning.

\end{abstract}

\input{texs/intro.tex}   	
\input{texs/problem.tex}   	
\input{texs/algorithm.tex}  
\input{texs/closedAlg-work.tex}  
\input{texs/experiments.tex}
\input{texs/related.tex} 	
\input{texs/conclusion.tex} 

\bibliographystyle{abbrv}

\begin{small}
\small{\bibliography{mybibJ}}
\end{small}

\input{texs/appendix.tex} 

\end{document}

%% file: texs/intro.tex
\section{Introduction}
\label{sec:intro}

 An important task in the analysis of the voluminous loosely structured datasets of big data applications is the mining of frequent patterns. Mining frequent patterns is also a core process of other important data mining processes like classification, clustering  and  detection of outliers and it has applications in numerous practical domains like database design, query optimization, chemical compound prediction, protein-protein interaction etc. The number of patterns produced from pattern mining is often huge. Besides the efficiency, this can also reduce the effectiveness of the mining process, since it is not realistic to comprehend and store such a large number of patterns. It is well known that the real problem in pattern mining is not the efficiency of the mining process but rather the usability of the mined frequent pattern set. One way to address this problem is to “summarize” the patterns. A high-quality pattern summarization technique essentially preserves the general information on the frequent patterns while reducing substantially the number of patterns.

\vspace*{.5ex}\noindent{\bf The framework.} In this paper we study summarizing patterns extracted from large tree data. A tree is a data structure most frequently used in big data senarios. Trees organize the data in the form of hierarchies, but at the same time they have the flexibility of allowing arbitrary horizontal and vertical expansions. Trees are employed in data models like XML and JSON which are standard formats for  integrating, exchanging, exporting and representing web data and in databases including (e.g., in the NoSQL database MongoDB).

Tree pattern mining has been researched extensively, because it is important for data analysis  \cite{AsaiAKASA02,TermierRS02,Nijssen03,AsaiAUN03,XiaoYLD03,WangHPZWS04,NijssenK04,LianMCY05,ChiXYM05,Zaki05,Zaki05tkde,HidoK05,GoethalsHB05,TatikondaPK06,TermierRSOWM08,TanHDCF08,DriesN12,KibriyaR13,ZhangDW15,WuT15,WuT16,WuTSBDR18}.
Two parameters characterize the tree patterns mined from the data: (a) the kind of morphism used to match the nodes of a tree pattern to the nodes of the data structure, and (b) the kind of edges the tree pattern can have. Edges can be either child edges or descendant edges. A child edge matches an edge in the data structure. A descendant edge matches a path in the data structure. In previous works, the morphisms adopted are almost exclusively embeddings. Embeddings are injections which preserve node labels and not-on-the-same-path relationships between the pattern nodes \cite{Zaki05,Zaki05tkde}. When an embedding matches the edges of the tree pattern  to edges in the data tree (that is, the edges of the pattern are seen as child edges), it is an {\em isomorphism} \cite{ChiXYM05,KibriyaR13}.

Initially, the extracted tree patterns were {\em induced} patterns  \cite{AsaiAKASA02,Nijssen03,AsaiAUN03,XiaoYLD03,NijssenK04,WangHPZWS04,ChiXYM05,HidoK05,GoethalsHB05,TermierRSOWM08,DriesN12,KibriyaR13}. Induced patterns are matched to the data structure through isomorphisms  and, of course, they contain exclusively child edges. Gradually,  the mined tree patterns evolved to {\em embedded} patterns \cite{Zaki05,Zaki05tkde,TatikondaPK06,TanHDCF08,ZhangDW15,WuT15,WuTSBDR18}. Embedded patterns generalize induced patterns since they are matched to the data structure through embeddings, and contain exclusively descendant edges.
In this sense, embedded patterns are able to discover relationships ``hidden'' (or embedded) deeply within the data structure which might not be discovered by induced patterns \cite{Zaki05tkde,Zaki05}.  However, extracting embedded patterns is computationally more challenging than extracting induced patterns. For instance, the problem of checking whether an isomorphism of an unordered tree pattern with child edges exists in a tree is polynomial \cite{Kilpel92treematching}. In contrast, the problem of checking whether an embedding of an unordered tree pattern with descendant edges exists in a tree is NP-Complete  \cite{KilpelainenM95}.

Current algorithms that mine trees almost exclusively focus on extracting patterns from a collection of small trees. Extracting patterns from a large tree was addressed only recently \cite{WuT15,WuT2016Dse,WuTSBDR18}.  The large tree framework generalizes the framework of a set of small trees. Indeed, a set of small trees can be represented by a single large tree with an unlabeled virtual root.  A plethora of applications input and generate datasets which are structured as a large data tree \cite{ChoiLL14,WangC17}.

The present study is placed in the broad framework of mining {\em embedded} patterns from {\em large tree data} which can be a list of small trees or a single large tree.

\vspace*{.5ex}\noindent{\bf The problem.} Typically, with a high frequency threshold, mining will generate only trivial patterns, while with a low frequency threshold, it will generate a huge number of patterns. This drawback seriously impedes the utility and usage of frequent-pattern mining. To address this issue, an obvious solution would be to summarize the patterns. 
A well-known solution for high-quality summarization consists in generating clusters of frequent patterns based on a similarity metric, and then choose and return only a representative pattern from each cluster \cite{XinHYC05}.  This reduction enjoys many advantages: (a)  Computing a much smaller pattern set results in improved efficiency; (b) a small number of mined patterns allows the users to examine them and understand the mining results; and (c) larger and more complex datasets can be handled. In fact, in multiple real-world scenarios, the primary goal is not the enumeration of all frequent patterns. Rather, the extracted patterns are potentially exploited as input in the following analysis or modeling phases, and therefore, a small set of representative patterns might be sufficient \cite{asanCSBZ07}.

A popular summarization technique employed in the setting of induced patterns mined from multiple small data trees is to compute maximal and closed patterns. A frequent pattern is closed if there is no isomorphic pattern with the same support. A maximal pattern is a frequent pattern such that there is no frequent isomorphic proper superpattern. 
When mining patterns from multiple small trees, pattern support is the number of small trees containing the pattern. This way of measuring pattern support is called {\em document support}. Unfortunately, these definitions do not apply to the broad setting we consider in this paper because: (a)~document support does not make sense in the setting of a large tree dataset, and (b)~the concept of isomorphic subpattern/superpattern is not meaningful in the setting of embedded patterns which involve embeddings and contain descendant edges. To the best of our knowledge, there is not any work on maximal or closed embedded tree patterns, even in the framework of multiple small trees.

\begin{figure}[!t]
    \centering%
     \scalebox{1}{ \epsfig{file=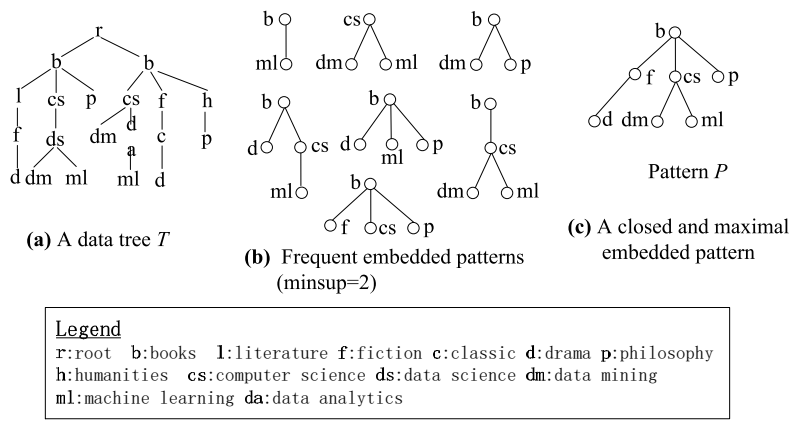}}%
     \caption{A data tree and frequent embedded patterns.}
      \label{intro}
\end{figure}

Let's look at the data tree of Figure \ref{intro}(a) which integrates categorized bibliographic information from two different data sources (the book entries are omitted). We adopt root support for the patterns (the number of the distinct images of the root of the pattern under all possible embeddings of the pattern to the tree dataset) and the support threshold is 2. Figures \ref{intro}(b) and (c) show some frequent embedded patterns. As one can see, pattern $P$ of Figure \ref{intro}(c) is larger in size and more informative than the other patterns.  Instead of presenting to the user all these patterns, we can just present the pattern $P$. The patterns of Figure \ref{intro}(b) might not be isomoprphic subpatterns of  pattern $P$ but they are all emdedded subpatterns of $P$ (that is, they have an embedding to $P$). Later, we introduce the concepts of embedded maximal and embedded closed pattern which can be used for clustering the embedded frequent patterns and also can act as representatives of the embedded frequent pattern clusters. Based on these definitions, it will become evident that pattern $P$ is a closed (and a maximal) embedded frequent pattern. This fact justifies omitting the patterns of figure  \ref{intro}(b) on behalf of pattern $P$.

In this paper, we address the problem of summarizing embedded frequent unordered patterns mined from a single or multiple large data trees by defining the concepts of maximal and closed embedded tree patterns and by designing algorithms for mining those types of patterns.

\vspace*{.5ex}
\noindent{\bf Contribution.} The main contributions of this paper are the following:
:
\begin{list}{}{\setlength{\leftmargin}{\parindent}\setlength{\parsep}{0cm}%
\setlength{\partopsep}{0cm}\setlength{\itemsep}{0cm}\setlength{\parskip}{0cm}%
\setlength{\labelwidth}{\parindent}%
\setlength{\topsep}{0cm}}

\item[$\bullet$] Defining meaningful closed and maximal patterns is tricky in the context of embedded patterns over a single large tree. Therefore, we employ a new way for defining pattern support which enjoys useful monotonic characteristics. Further, in order to account for embedded patterns, we generalize the concept of subpattern by introducing the concept of embedded subpattern.  Based on these concepts, we introduce closed and maximal embedded unordered tree patterns over large  tree data and study their properties.
\item[$\bullet$] We present a frequent pattern mining algorithm extended with a local closedness checking technique, called {\em closedEmbTM-eager}, to eagerly eliminate non-closed patterns. To allow for closedness checking, {\em closedEmbTM-eager} computes and stores the embedded occurrence lists of all the nodes of candidate frequent patterns. 
\item[$\bullet$] To further reduce the generation of frequent patterns which turn out to be non-closed, we design pruning rules to proactively detect branches in the pattern search space that do not correspond to closed patterns. These branches are pruned to avoid unnecessary computations. The pruning rules are leveraged in the design of a new algorithm,  named {\em closedEmbTM-prune}, for mining all the closed and maximal embedded frequent patterns from large data trees.
\item[$\bullet$] We conducted a comprehensive experimentation to assess the scalability and performance of our algorithms on synthetic and real datasets. We also compared with a post-processing algorithm which first extracts all frequent patterns using a state-of-the-art algorithm \cite{WuTSBDR18} before filtering out non-closed (or non-maximal) ones. The experimental results show that: 
    (a) Our algorithms generate a complete closed and maximal pattern set which is substantially smaller than the set of embedded patterns generated by an embedded pattern miner.
    (b)  Algorithm {\em closedEmbTM-eager} outperform the baseline post-processing algorithm as the latter checks for closedness the entire frequent pattern set; this set might be exponentially larger than the closed pattern set.
    (c) On sparse datasets, {\em closedEmbTM-prune} has comparable time and memory performance with {\em closedEmbTM-eager}; this is due to a small overhead incurred by the application of the pattern pruning rule;
    (d)~On dense datasets, {\em closedEmbTM-prune} outperforms in all cases {\em closedEmbTM-eager} in regard to run-time by orders of magnitude and also in regard to memory footprint and scalability.
   \end{list}

\vspace*{.5ex}\noindent\textbf{Paper outline.} The following section introduces the concept of embedded subattern, formally describes the problem addressed and provides preliminary results. Section \ref{sec:prelim} sets the framework of  our approach for extracting closed embedded patterns.  Algorithms {\em closedEmbTM-eager} and {\em closedEmbTM-prune} are presented in Sections \ref{sec:eagerAlgo} and \ref{sec:algoPrune}, respectively.  In Section \ref{sec:experiments}, we present and analyze our comparative experimental results. We review related work in Section \ref{sec:related}. A conclusion and suggestions for future work are provided in Section Section \ref{sec:conclusion}.

%% file: texs/problem.tex
\section{Framework, Preliminary Results and Problem Statement}
\label{sec:problem}

\subsection{Generalities and Support Definition}
\label{subsec:general}
\noindent\textbf{Trees.} A {\em rooted labeled tree} is a directed acyclic connected graph $T=(V,E)$ where $V$ is a set of nodes and $E\subseteq V \times V$ is a set of edges satisfying the following properties:  (1)~There is a function $lb$, called labeling function, which maps nodes to labels. (2)~Set $V$ has a unique node called the {\em root} which has no incoming edges; (3)~Every node in $V$ is linked to the root through a unique path. A tree is {\em ordered} if it there is a predefined left-to-right order among the child nodes of each node. Otherwise, it is {\em unordered}. We define the {\em size} of a tree to be the number of its nodes. Unless stated differently, in the following trees are rooted, labeled, unordered trees. Table \ref{tab:notation} summarizes the notation used throughout this paper.

\begin{table}[!h]
\footnotesize
      \center
      \begin{tabular}[b]{|l|l|}
            \hline
                    Notation	& Description\\
            \hline
             \hline
             $T$    & data tree \\
            lb(.)	& label function on graph nodes	\\	
               $P' \sqsubseteq_i P$	& pattern $P'$ is an isomorphic subpattern of pattern $P.$ \\
               $P' \sqsubseteq_e P$	& pattern $P'$ is an embedded subpattern of pattern $P.$ \\
                $L_{root}(P_1|P_2)$	& the set of the images of nodes $n_1, \ldots, n_k$ under all embedding of $P_2$ to $T$, where\\
            &   $n_1, \ldots, n_k$   are are images (nodes in $P_2$) of the root of $P_1$ under all embeddings\\
            & of $P_1$ to $P_2$. 	\\	
             {\em rml}	& rightmost leaf of a given pattern \\
             {\em k-pattern} 	& a pattern of size $k$\\
               $[P]$ 	& equivalence class of pattern $P$\\
               $P_{x}^{i}$	& the pattern formed by adding a child node labeled by $x$ to the node with position\\
               & $i$ in $P$ as its {\em rml}. \\
             $P_{x}^{i} \otimes_c P_{y}^{j}$	& child join of $P_{x}^{i}$ and $P_{y}^{j}$	\\	
             $P_{x}^{i} \otimes_s P_{y}^{j}$	& cousin join of $P_{x}^{i}$ and $P_{y}^{j}$	\\	
           $A//B$	&  node $B$ is a descendant of node $A$ 	\\	
              $OC(P)$ 	&  occurrence relation of pattern $P$ on $T$  \\	
        $L_X$ & occurrence list of pattern node $X$ on $T$  	\\	
        $OL(P)$	&  occurrence list set of $P$: the set of all the occurrence lists of the nodes of $P$ on $T$  \\	
         $P_1 \leq P_2$	& order relation on patterns defined on page 18.	\\	
          $P \equiv P'$	& patterns $P$ and $P'$ are occurrence equivalent.	\\

            \hline

        \end{tabular}
   \caption{Notation used throughout the paper.}
    \label{tab:notation}
  \vspace*{-2ex}
\end{table}

\begin{figure}[!t]
    \centering%
     \scalebox{1}{ \epsfig{file=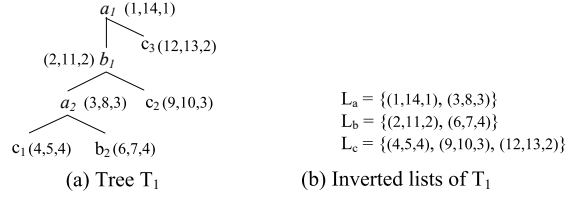}}%
     \caption{A tree and its inverted lists.}
      \label{fig:treewithInvLsts}
\end{figure}

\vspace*{.5ex}\noindent\textbf{An encoding scheme for data trees.} We employ the regional encoding scheme \cite{bruno02} to encode the input data tree $T$ in a preprocessing step. This encoding represents each node in $T$ by a triple ($begin$, $end$, $level$) whose elements correspond, respectively, to the  first, and last encounter of the node in a depth-first preorder traversal of $T$, and its level in the tree.
For every label $a$ in $T$, the pre-processing step produces an inverted list $L_a$ for $a$. List $L_a$ comprises the triples of the nodes in $T$ labeled by $a$, ordered on their $begin$ field. In Fig. \ref{fig:treewithInvLsts}(a) one can see a data tree $T_1$ and the encoding of the nodes. Nodes with the same label in $T$ are distinguished using different subscripts (e.g., nodes $a_1$ and $a_2$). In Fig. \ref{fig:treewithInvLsts}(b) one can see the inverted lists of the labels of the tree.

\vspace*{.5ex}\noindent\textbf{Tree patterns.} The tree patterns considered in the literature for mining fall into two categories: patterns with {\em child edges} (that is edges representing child relationships) and patterns with {\em descendant edges} (that is edges representing descendant relationships). Patterns with descendant edges are more general since a child relationship between two nodes implies also a descendant relationship. The opposite is not true. The focus in this paper is on patterns with descendant edges which are called embedded patterns and are defined below.

\vspace*{1ex}\noindent\textbf{Tree morphisms.} In order to decide if a pattern occurs in a data tree different types of morphisms are adopted in the literature. Given a pattern $P$ and a tree $T$, a {\em homomorphism} from a pattern $P$ to a data tree $T$ is a mapping $m$ associating every node of $P$ to a node in $T$, such that: (1)~for every node $x\in P$, $lb(x)$ = $lb(m(x))$; and (2) for every edge ($x,y$) $ \in P$, if ($x,y$) is a child edge, ($m(x),m(y)$) is an edge of $T$, while if ($x,y$) is a descendant edge, $m(y)$ is a descendant of $m(x)$ in $T$.

\begin{figure}[!t]
    \centering%
     \scalebox{1}{ \epsfig{file=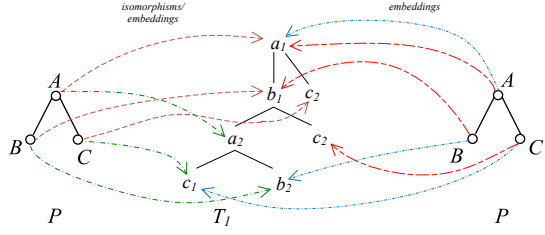}}%
     \caption{Embeddings of a pattern $P$ to the data tree $T_1$ of Figure \ref{fig:treewithInvLsts}(a)}
      \label{fig:morphisms}
\end{figure}

Different constraint versions of homomorphisms have been considered in previous mining papers: an {\em isomorphism} from pattern $P$ to $T$ is an injective mapping $m$ associating every node of $P$ to a node in $T$, such that: (1)~for every node $x$ of $P$, $lb(x)$ = $lb(m(x))$; and (2)~($x,y$) is an edge of $P$ {\em if and only if} ($m(x),m(y)$) is an edge in $T$. In Fig. \ref{fig:morphisms} one can see the two possible isomorphisms of pattern $P$ to the data tree $T_1$ shown in Fig. \ref{fig:treewithInvLsts}(a). Obviously, isomorphisms are special cases of homomorphisms. The  patterns mined when isomorphisms  are adopted for mapping the patterns to the data tree are qualified as {\em induced} \cite{AsaiAKASA02,Nijssen03,ChiXYM05}.

Embeddings are another kind of restricted homomorphisms considered in the literature: consider a pattern $P$ which has descendant edges. An {\em embedding} from $P$ to $T$ is an injective function $m$ associating nodes in $P$ to nodes in $T$, such that: (1)~for every node $x\in P$, $lb(x)$ = $lb(m(x))$; and (2)~$(x,y)$ is an edge in $P$ {\em iff} $m(y)$ is a descendant of $m(x)$ in $T$. Consequently, if $x_1, \ldots, x_n$ are the child nodes of a node $x$ in $P$,
$m(x_i)$ and $m(x_j),$ $i, j \in [1,n],\;  i\neq j,$ are not on the same path (or coincide) in $T$. We term this constraint satisfied by embeddings {\em sibling constraint}. Obviously, an embedding is also a restricted case of a homomorphism. In Fig. \ref{fig:morphisms} one can see the data tree $T_1$ of Fig. \ref{fig:treewithInvLsts}(a), a pattern $P$ and four embeddings of $P$ to $T_1$. Pattern $P$ has one more embedding to $T_1$ which is not shown in the figure. Among the embeddings shown on the figure, the two on the left are isomorphisms. When embeddings are used, the mined patterns are called {\em embedded} \cite{Zaki05,Zaki05tkde}. In contrast to an isomorphism, an embedding can associate two adjacent nodes in the pattern to non-adjacent nodes on a path in the data tree. 

The set of frequent induced patterns whose edges are seen as descendant edges is a subset of the set of frequent embedded patterns on the same tree $T$. In this paper, the mined patterns are {\em embedded patterns}.

\vspace*{1ex}\noindent\textbf{Support.} When the dataset is a set of small trees, a natural way to define the support (frequency) of a pattern is to let it be equal to the number of trees that contain the pattern (called in the literature {\em document frequency}).  However, document frequency is meaningless in the framework of a dataset which is a single large tree, as there is only one document. Therefore, a new definition for pattern support is needed for the setting of a single large tree. A desirable property for the support is the antimonotonicity property \cite{Zaki05tkde}. The antimonotonicity property is fundamental for efficiently enumerating  patterns in the mining process. 

In this framework, a natural candidate for the definition of pattern support is the number of occurrences (that is, embeddings) of the pattern in the dataset ({\em occurrence frequency}). Unforunately, occurrence frequency does not enjoy the antimonotonicity property. As we want to take advantage of the benefits of antimonotonicity, we adopt {\em  root frequency} for embedded tree patterns:

\begin{definition}[Root frequency]
Let $R$ be the root of a tree pattern $P$ on a data tree $T$. The {\em root frequency} of $P$ on $T$ is the number of distinct images (which are nodes in $T$) of $R$ under all possible embeddings of $P$ to $T$.
\end{definition}

It can be shown \cite{WuT15} that root frequency has the antimonotonicity property.

\subsection{Embedded Subpatterns}

\begin{figure*}[!t]
    \centering%
     \scalebox{1}{ \epsfig{file=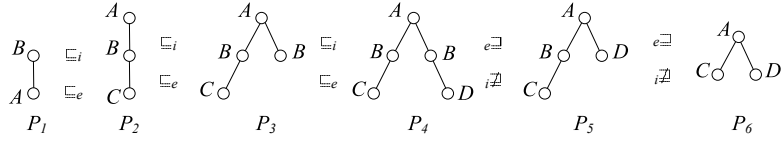}}%
     \caption{Tree patterns.}
      \label{fig:EmbedSubPatterns}
\end{figure*}

Usually, the mining process on a large data tree results in a large number of embedded tree patterns. It is therefore desirable to devise a concise and nonredundant representation for frequent embedded tree patterns. To this end, we introduce later the concepts of {\em closed} and {\em maximal} frequent embedded tree pattern. Closed and maximal tree patterns have been introduced before albeit in the context of induced tree patterns \cite{ChiXYM05}. To define closed tree patterns we need the concept of {\em subpattern}. In the case of induced patterns, isomorphic subpatterns are employed: a pattern $P'$ is an {\em isomorphic subpattern} of a pattern $P$ iff there is an isomorphism from $P'$ to $P$. This means that $P'$ is a subtree of $P$. In other words, $P'$ can be obtained from $P$ by removing, in sequence, zero or more dangling edges (that is, edges incident to a leaf or root node) from $P$. We write $P' \sqsubseteq_i P$ to indicate that $P'$ is an isomorphic subpattern of $P$. Consider, for instance, the patterns shown in Figure \ref{fig:EmbedSubPatterns}. As one can see, $P_1 \sqsubseteq_i  P_2 \sqsubseteq_i  P_3 \sqsubseteq_i  P_4$, while $P_5 \not\sqsubseteq_i  P_4$, $P_6 \not\sqsubseteq_i  P_4$ and $P_6 \not\sqsubseteq_i  P_5$. In particular, a {\em root isomorphic subpattern} is a special type of isomorphic subpattern: a pattern $P'$ is a {\em root isomorphic subpattern} of a pattern $P$ if and only if there is an isomorphism from $P'$ to $P$ which maps the root of $P'$ to the root of $P$. For example, the pattern $P_2$ of Figure \ref{fig:EmbedSubPatterns} is a root isomorphic subpattern of $P_3$, while $P_1$ is an isomorphic subpattern but not root isomorphic subpattern of $P_2$.

However, the concept of isomorphic subpattern is not sufficient for embedded patterns since it cannot account for descendant relationships between nodes. We therefore generalize the concept of subpattern by introducing the concept of embedded subpattern to account for embedded patterns.

\begin{definition}[Embedded subpattern]
A pattern $P'$ is an {\em embedded subpattern} of a pattern $P$ (and $P$ is an embedded superpattern of $P'$) iff there is an embedding from $P'$ to $P$. We write $P' \sqsubseteq_e P$ to indicate that $P'$ is an {\em embedded subpattern} of $P$, .
\end{definition}

Consider again the patterns of Figure \ref{fig:EmbedSubPatterns}. One can see that $P_1 \sqsubseteq_e  P_2 \sqsubseteq_e  P_3 \sqsubseteq_e  P_4$ and also that $P_6 \sqsubseteq_e  P_5 \sqsubseteq_e  P_4$. Besides removing dangling edges, as is the case with isomorphic subpatterns, an embedded subpattern of a pattern $P$, can be obtained from $P$ by replacing paths in $P$ by edges. For instance, $P_6$ which is an embedded subpattern of $P_4$, can be obtained from $P_4$ by replacing the paths from $A$ to $C$ and from $A$ to $D$  by the edges $A/C$ and $A/D$, respectively.

Embedded subpatterns are related to isomorphic subpatterns the way the next proposition shows.

\begin{proposition}
$P' \sqsubseteq_e P$ if $P' \sqsubseteq_i P$. The opposite is not necessarily true.
\end{proposition}

Indeed, if $P' \sqsubseteq_i P$, there is an isomorphism from $P'$ to $P$. Therefore, there is an embedding from $P'$ to $P$, i.e., $P' \sqsubseteq_e P$. As we can see in Figure \ref{fig:EmbedSubPatterns}, all the induced subpatterns are also embedded subpatterns. There are also examples in the figure of a pattern which is an embedded  subpattern of another pattern but not an isomorphic subpattern. For instance, $P_5 \sqsubseteq_e  P_4$ and $P_5 \not\sqsubseteq_i  P_4$.

\subsection{ Maximal and Closed Embedded Patterns and Problem Definition}
Frequent induced patterns over multiple trees with document support are summarized using closed and maximal (induced) patterns. We want to introduce analogous concepts for embedded frequent patterns over a single tree with root support.

In the context of induced patterns over multiple trees with document support, a closed pattern is a frequent pattern such that no proper isomorphic superpattern 
has the same support. Note that an isomorphic superpattern of a pattern cannot have higher document support than the pattern itself. As a consequence, an induced pattern $P$ is closed iff there is a proper superpattern $P'$ which has an isomorphism $i'$ to $T$ whenever $P$ has an isomorphism $i$ to $T$ such that $i'$ is an extension of $i$. In the context of embedded patterns we use the following definition for closed patterns:

\begin{definition}[Closed pattern]
Given a data tree $T$, an embedded frequent pattern $P$ is {\em closed} iff there is no proper embedded superpattern $P'$ satisfying the following property: for every embedding $e$ of $P$ to $T$ there is also an embedding $e'$ of $P'$ to $T$ such that $e(root(P)) = e'(m(root(P)))$ for some embedding $m$ of $P$ to $P'$.
\label{def:close}
\end{definition}

\begin{figure}[!t]
    \centering%
     \scalebox{0.9}{ \epsfig{file=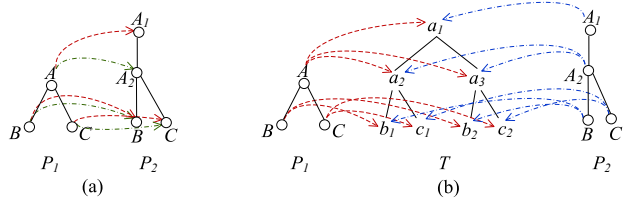}}%
     \caption{An example for embedded pattern closedness and maximality definition.}
 \label{fig:closedDefn}
\end{figure}

Consider, for instance, the patterns $P_1$ and $P_2$ and  the data tree $T$ depicted in Figure \ref{fig:closedDefn}. Pattern $P_1$ is an embedded subpattern of $P_2$ as there is an embedding from $P_1$ to $P_2$. In fact, there are two such embeddings, shown in Figure \ref{fig:closedDefn}(a), which map the root $A$ of $P_1$ to the nodes $A_1$ and $A_2$ of $P_2$. Pattern $P_1$ has six embeddings to $T$ which map the root of $P_1$ to node $a_2$ in $T$ (one embedding), to $a_3$  (one embedding) and to $a_1$ (four embeddings). In Figure \ref{fig:closedDefn}(b), the dashed lines show possible mappings of the nodes of $P_1$ and $P_2$ to $T$. Pattern $P_2$ has two embeddings to $T$ which map the root of $P_2$ to $a_1$. One can see that for every one of the six embeddings of $P_1$ to $T$, one of the two embeddings of $P_2$ is its extension. For instance, for the embedding $e$ of $P_1$ to $T$ ($e(A) = a_1$, $e(B) = b_1$,  $e(C) = c_1$), the embedding $e'$ of $P_2$ to $T$ ($e'(A_1) = a_1$, $e'(A_2) = a_1$, $e'(B) = b_1$, $e'(C) = c_1$)  is an extension of $e$ based on the embedding $m(A) = A_1$, $m(B) = B$, and $m(C) = C$ of $P_1$ to $P_2$. Therefore, $P_1$ is not closed, and $P_2$ or another superpattern which will turn out to be closed should be returned to the user instead. One can see that, in general, if a pattern $P$ has an embedding to a pattern $P'$ which maps the root of $P$ to the root of $P'$, $P$ is not closed.

Note that it would not be appropriate to define closed patterns in a simpler way based on root support (as this is done with induced patterns and document support). For instance, consider the data tree $T$ and the patterns $P_1$ and $P_2$ of Figure \ref{fig:closedDefn}. Assume that the support threshold is 1. As mentioned above, $P_1 \sqsubseteq_e P_2$. The root support of $P_1$ on $T$ is 3 while that of $P_2$ is 1. Therefore, if pattern closedness is decided based on root support, the presence of the embedded superpattern $P_2$  does not allow us to deduce, as Definition \ref{def:close} postulates, that $P_1$ is not closed. In fact, it can be shown that, based on root support, $P_1$ is closed with support threshold 3, which counters our intuition.

We provide next a proposition which characterizes closed embedded patterns. Let $L_{root}(P_1)$ denote the set of the images of the root of pattern $P_1$ under all embeddings of $P_1$ to a data tree $T$. Given two patterns $P_1$ and $P_2$ such that $P_1 \sqsubseteq_e P_2$, let $n_1, \ldots, n_k$ be the nodes in $P_2$ which are images of the root of $P_1$ under all embeddings of $P_1$ to $P_2$. Then, let $L_{root}(P_1|P_2)$ denote the set of the images of nodes $n_1, \ldots, n_k$ under all embedding of $P_2$ to $T$. As an example, in Figure \ref{fig:closedDefn}, $L_{root}(P_1|P_2) = \{a_1, a_2, a_3\}$.

\begin{proposition}
A frequent pattern $P$ is closed if there is no proper embedded superpattern $P'$ such that $L_{root}(P) = L_{root}(P|P')$
\label{prop:close}
\end{proposition}

The proof follows easily from the definitions. In the example of Figure \ref{fig:closedDefn}, $P_1$ is not closed since for its proper embedded superpattern $P_2$, $ L_{root}(P_1) =$  $L_{root}(P_1|P_2) = \{a_1, a_2, a_3\}$.

Maximal patterns can be defined without referring explicitly to the embeddings to the data tree (only their number matters):

\begin{definition}[Maximal pattern]
Given a data tree, an embedded frequent pattern is {\em maximal} if and only if there is no proper embedded superpattern which is also frequent.
\label{def:max}
\end{definition}

Clearly, every maximal pattern is also closed. The opposite is not necessarily true. That is, the set of closed patterns is a superset of the set of maximal patterns for a given frequency threshold. Although maximal patterns provide a tighter summarization of the frequent embedded patterns than closed patterns, closed patterns are important as are they are more informative than maximal patterns about the frequent embedded pattern set they summarize.

\vspace*{1ex}\noindent\textbf{Problem statement.} Given a minimum support threshold and a large data tree $T$, find all the frequent unordered maximal and all the frequent unordered closed embedded patterns in $T$.

One of the reasons this problem is challenging is that, unlike the property of a pattern being frequent, pattern closedness (resp. maximality) does not have the antimonotonicity property: a superpattern of a non-closed  (resp. non-maximal) pattern can be closed (resp. maximal).

%% file: texs/algorithm.tex
\section{Mining Framework}
\label{sec:prelim}

We present in this section the framework of our approach for mining frequent embedded patterns: we introduce the techniques for generating candidate patterns and those for computing their support.

\subsection{Candidate Generation}
\label{subsec:candGen}

In order to mine embedded patterns, our approach alternates between a step that generates candidate patterns and a step that computes the support of the patterns.  The first step employs a process for systematically generating potentially frequent candidate patterns. For the second step, a novel technique is designed which computes the support of candidate patterns in an incremental way. We start by outlining below the candidate generation. 

\vspace*{1ex}\noindent\textbf{Canonical form.} Because unordered patterns can have many different isomorphic representations, an efficient pattern mining process needs a method for avoiding the superfluous generation of isomorphic pattern representations and the superfluous isomorphism checking between pattern representations. 

Several mining algorithms exploit the notion of tree canonical form which is a unique ordered representation of an unordered pattern and can be used as a representative of this pattern. Here we use a canonical from similar to the one adopted by \cite{Zaki05} for tree patterns. A variety of alternative  canonical tree representations is studied in \cite{ChiYM05}.

\vspace*{1ex}\noindent\textbf{Equivalence classes.} For the generation of candidate patterns we follow the equivalence class-based pattern technique presented in \cite{Zaki05tkde,Zaki05}. We briefly outline next this technique, in order to also introduce concepts and notation which are needed along the paper.

The nodes of a pattern $P$ are encoded by their {\em depth-first position} in $P$. The depth-first position of a node is the number assigned to the node in its first encounter during a depth first traversal of the pattern tree which sequentially assigns numbers starting with zero. The {\em rightmost leaf} of $P$ is the node with the largest depth-first position and is denoted as {\em rml}. The {\em rightmost path} of $P$ is the path from the root of $P$ to the node {\em rml}. The {\em immediate prefix} of $P$ is the sub-pattern of $P$ obtained by removing the node {\em rml} from $P$. A {\em prefix} of $P$ is the sub-pattern of $P$ obtained by a series of deletions of the node {\em rml} .

We call {\em k-pattern} a pattern with $k$ nodes. The {\em equivalence class} of a (k-1)-pattern $P$ (denoted as $[P]$) is the set of all frequent k-patterns whose immediate prefix is $P$. The $k$-pattern constructed by adding a node labeled by $x$ to the node at position $i$ in $P$ as the rightmost leaf node of $[P]$ is denoted by $P_{x}^{i}$.

Let the term {\em k-pattern} refer to a pattern of size $k$. Given a (k-1)-pattern $P$ ($k\geq$ 1), its {\em equivalence class} is the set of all frequent k-patterns which have $P$ as immediate prefix. We write $P_{x}^{i}$ to denote a $k$-pattern in $[P]$ formed by adding a child node labeled by $x$ to the node with position $i$ in $P$ as its rightmost leaf node.

\begin{figure}[!t]
    \centering%
     \scalebox{0.8}{ \epsfig{file=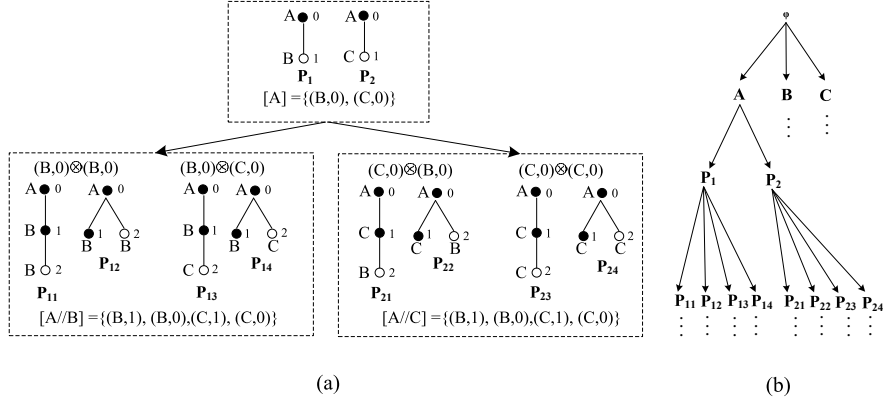}}%
     \caption{Equivalence class expansion example. The immediate prefix of a pattern is represented by its black nodes. }
 \label{fig:patternExpan}
\end{figure}

\vspace*{1ex}\noindent\textbf{Expansion of patterns.} The successor equivalence classes of an equivalence class $[P]$ are constructed by expanding the patterns in $[P]$.  This expansion works as follows: each pattern $P_{x}^{i} \in [P]$ is {\em joined} with any other pattern $P_{y}^{j} \in [P]$, including itself (self expansion). The result is the equivalence class $[P_{x}^{i}]$. Two types of expansions are used: {\em child expansion}, implemented by a join operation, denoted $P_{x}^{i} \otimes_c P_{y}^{j}$ ({\em child join}), and {\em cousin expansion}, implemented by a join operation, denoted $P_{x}^{i} \otimes_s P_{y}^{j}$ ({\em cousin join}).
Formally:
\begin{list}{}{\setlength{\leftmargin}{\parindent}\setlength{\parsep}{0cm}%
\setlength{\partopsep}{0cm}\setlength{\itemsep}{0cm}\setlength{\parskip}{0cm}%
\setlength{\labelwidth}{\parindent}%
\setlength{\topsep}{0cm}}
\item[$\bullet$] {\em Child join} can be applied only if $j=i$, and $P_{x}^{i} \otimes_c P_{y}^{j} = Q_{y}^{k-1}$, where  $Q = P_{x}^{i}$. Note that $k-1$ is the position of the last element, $x$, in the k-pattern $P_{x}^{i}$ since the numbering of nodes starts from 0.
\item[$\bullet$] {\em Cousin join }can be applied only if $j \leq i$, and $P_{x}^{i} \otimes_s P_{y}^{j} =  Q_{y}^{j}$, where  $Q = P_{x}^{i}$.
\end{list}

No expansion is possible if $i<j$. The two outcome patterns $Q_{y}^{k-1}$ and $Q_{y}^{j}$ are respectively called {\em child} and {\em cousin} expansions of $P_{x}^{i}$ by $P_{y}^{j}$. Joining $P_{x}^{i}$ with all elements $P_{y}^{j}$ of $[P]$ produces all possible $k$-patterns in class $[P_{x}^{i}]$. The patterns $P_{x}^{i}$ and $P_{y}^{j}$ are called, respectively, the {\em left-parent} and {\em right-parent} of a join outcome, respectively. Notice that each join outcome has its left-parent as its immediate prefix, and its righmost leaf is the rightmost leaf of its right-parent. 

Fig.~\ref{fig:patternExpan} shows an example of an equivalence class expansion. The depth-first position of each node is shown by the node in the figure.

An example of the equivalence class-based pattern expansion is given in Fig.~\ref{fig:patternExpan}. The number associated with each pattern node in the figure is the depth-first position of that node. The left and right child rectangles surround, respectively, the patterns of the equivalence classes of the patterns $A//B$ and $A//C$ (note that $A//B$ denotes that node $A$ is an ancestor of node $B$).

\vspace*{1ex}\noindent\textbf{Pattern search tree.} Starting with patterns of size $k$=1, we can iteratively apply the equivalence class expansion to produce larger patterns. The construction of a pattern from another pattern defines a partial order, $\prec$, on patterns. This partial order is characterized by the prefix containment relationship, that is, $P \prec Q$ iff $P$ is a prefix of $Q$. It is represented as a {\em pattern search tree}, in which node $Q$ is descendant of $P$ iff $P \prec Q$. All the children of a given node $P$ belong to its equivalence class $[P]$, since they all have the same immediate prefix $P$.  Patterns in the search tree are organized in layers, where the 0th layer is the root denoting an empty pattern,  and the $k$th layer consists of patterns with $k$ nodes. Fig. \ref{fig:patternExpan}(b) shows a fragment of an example pattern search tree for the patterns of Fig. \ref{fig:patternExpan}(a). 

The internal nodes of the search tree represent frequent patterns in canonical form. The execution of the pattern mining algorithm follows the growth a pattern search tree from the root. 
One-node frequent patterns are used as the initial pattern seeds. To augment a node $P_{x}^{i}\in [P]$, we require that $P_{x}^{i}$ is frequent and in canonical form. Such a pattern $P_{x}^{i}$ can only be expanded using patterns from $[P].$  To guarantee that all the children of $P_{x}^{i}$ in the tree will be generated, $[P]$ comprises all the possible frequent expansion outcomes of $P$  (including non-canonical). Since infrequent or non-canonical patterns are not further grown, they correspond to leaf nodes.

\subsection{Support Computation}
\label{subsec:supComp}

\vspace*{1ex}\noindent\textbf{From homomorphic occurrences to embedded ocurrences.}
We call {\em embedded occurrence} or simply {\em occurrence} of a pattern $P$ on a data tree $T$ a tuple indexed by the nodes of $P$ whose values are the images of the corresponding nodes in $P$ under an embedding of $P$ to $T$. The {\em occurrence relation} $OC(P)$ of $P$ on $T$ is the relation containing the occurrences of $P$ under all possible embeddings of $P$ to $T$. If $X$ is a node in $P$, the {\em occurrence list $L_X$ of $X$ on $T$} is the projection of relation $OC(P)$ on attribute $X$ sorted in the preorder appearance of the nodes in $T$. The {\em occurrence list set} $OL(P)$ of $P$ on $T$ is the set of all the occurrence lists of the nodes of $P$ on $T$; that is, $OL(P) =\{L_X \;|\;X\in nodes(P)\}$. The concepts of {\em homomorphic occurrence}, {\em homomorphic occurrence relation} $OC^h(P)$, {\em homomorphic occurrence list}, and {\em homomorhic occurrence list set} $OL^h(P)$ are defined analogously.

As mentioned above, in order to compute the support of a pattern $P$ on a data tree $T$, we need to compute the size of the embedded occurrence list $L_R$ of the root $R$ of $P$ on $T$. A straightforward method for computing $L_R$ consists of first computing $OC(P)$, and then projecting $OC(P)$ on column $R$ to get $L_R$. A majority of existing tree mining algorithms
\cite{AsaiAKASA02,AsaiAUN03,NijssenK04,ChiXYM05,Zaki05,Zaki05tkde,GoethalsHB05,TatikondaPK06,TermierRSOWM08,TanHDCF08,DriesN12,KibriyaR13} explicitly compute and store $OC(P)$.  Unfortunately, the problem of finding an unordered embedding of $P$ to $T$ is NP-Complete \cite{KilpelainenM95}. On the other hand, finding a homomorphism from $P$ to $T$ can be done in PTIME \cite{MiklauS04}.

Based on the above observation, we design a construction-pruning approach which first constructs $OC^h(P)$ and then, prunes non-embedded occurrences to get $OC(P)$ (as mentioned above an embedding is a special case of homomorphisms). For the first step (the homomorphic occurrence construction) we employ a holistic twig-join algorithm. The second step (the non-embedded occurrence filtering) is implemented by checking whether two sibling nodes of $P$ are mapped to nodes on the same path in $T$.

\vspace*{1ex}\noindent\textbf{Holistic Twig-join Algorithms for computing homomorphic occurrences.}
Holistic twig-join algorithms (e.g., $TwigStack$ \cite{bruno02}) are the state of the art algorithms for computing all the homomorphic occurrences of tree-patterns on tree data \cite{bruno02,chen06,GrimsmoBH10,LuLBW11,BacaKLL13}. These algorithms were initially developed for computing tree-pattern queries over trees. They can provably compute all the homomorphic occurrences in time linear on the size of the input (the data tree) and the output (set of homomorphic occurrences) \cite{bruno02}.

\vspace*{1ex}\noindent\textbf{Filtering out non-embedded occurrences.} Given a homomorphic occurrence $occ$ checking whether it is also an embedded one is performed by traversing, in post order, the nodes of $P$; if $Y_1, \ldots, Y_m$ are the child nodes of a node $X$ under consideration we check if for no $Y_i$ and $Y_j,\, i\neq j$, $occ.Y_i$ and $occ.Y_j$ are on one path (or coincide). As mentioned in Section \ref{subsec:general} this constraint is termed {\em sibling constraint}. Clearly, we can deduce that occurrence $occ$ is embedded if the sibling constraint is met. We can check whether two nodes are on one path in a tree $T$ efficiently: each node in $T$ is mapped to its positional representation which is a ($begin$, $end$, $level$) triple. This regional encoding allows for checking efficiently the existence of an descendant-ancestor relationship between two nodes: for two nodes $n_1$ and $n_2$, $n_2$ is a descendant of $n_1$ iff $n_2.begin>n_1.begin$, and $n_1.end>n_2.end$.

\vspace*{1ex}\noindent\textbf{Encoding pattern occurrences as occurrence list sets.}
For computing the pattern support we adopt an incremental technique which uses the materialized embedded occurrences of its parent patterns. We use an original materialization technique for the embedded occurrences of patterns which are processed earlier: we store the embedded occurrence lists of the nodes of the pattern instead of storing the embedded occurrence relation. The space efficiency of the method lies on the fact that the occurrence lists can encode in linear space in the number of pattern occurrences which is exponential on the number of pattern nodes \cite{bruno02}. In contrast, the state-of-the-art embedded pattern mining techniques \cite{Zaki05tkde,Zaki05} record information about all the embedded occurrences of the processed patterns in the data tree.

\vspace*{1ex}\noindent\textbf{Bitmap representation of occurrence lists.}
The pattern occurrence lists are stored as bitmaps. This choice allows for a big reduction in the storage cost. In addition, using bitmaps also allows for important CPU and I/O cost reductions. Indeed, occurrence list intersection of the nodes of a pattern can be realized as a two-step bitwise operation: the first step consists in bitwise ANDing the bitmaps of the occurrence lists of the pattern nodes. The second step materializes the intended occurrence list by fetching in memory, from the corresponding inverted list, the nodes indicated by the resulting bitmap. The benefit of the use of bitmaps in time consumption is twofold: certainly bitwise ANDing is cheaper than the intersection of the possibly long corresponding occurrence lists. Further, retrieving in memory only the nodes indicated by the result bitmap has less I/O cots than the retrieval of the entire---possibly very large---operand occurrence lists. The latter retrieval is necessary for applying the intersection operation in the traditional way. We used bitmaps for the storage and processing of occurrence lists initially in \cite{WuTW09,WuTWS13,WuTK14,WuT16J} in the context of evaluating queries using tree-pattern materialized views.

\section{An Eager Closedness Checking Algorithm}
\label{sec:eagerAlgo}

We provide in this section an algorithm which produces embedded frequent patterns and eagerly filters out non-closed ones. The pattern support is computed in this algorithm by extending a procedure presented in \cite{WuT15,WuTSBDR18} for computing frequent embedded patterns. The algorithm involves an efficient local closedness checking technique which is integrated in the mining process so that non-closed patterns can be eliminated early on.

\subsection{Closedness Checking }
\label{subsection:Closeness}

\vspace*{.5ex}\noindent\textbf{Pattern-subpattern relationship checking.} In order to determine if a given pattern $P$ is closed, we first need to check if $P$ is a proper subpattern of some other pattern $Q$, and find all the nodes in $Q$ which are images of the root of $P$ under an embedding from $P$ to $Q$ (Proposition \ref{prop:close}). For this, we can utilize the unordered tree inclusion algorithm described in \cite{KilpelainenM95}.  Provided two unordered trees $P$ and $Q$, this algorithm finds all the nodes $v$ in $Q$, such that there exists a root-preserving embedding from $P$ to the subtree of $Q$ rooted at node $v$. The time complexity of the algorithm is $O(|P|\times\ P_f \times 2^{2P_f}\times |Q|)$, with $P_f$ being the max outdegree of the nodes in $P$. However, in the experiments, we found that this algorithm runs very slowly for patterns with large fan-out. We therefore use a procedure we presented in \cite{WuT15} which computes embeddings of a pattern to a tree and this turns out to be faster.

Once we have obtained the root images $r_1, \ldots r_k$ of $P$ on $Q$, the computation of $L(P|Q)$ is straightforward, as it is the union of $L_{r_i}(Q)$s, $i \in [1,\ldots,k].$  Observe that, since the occurrence lists are represented as bitmaps, a bitwise OR operation will allow this union operation to be performed efficiently.

\vspace*{.5ex}\noindent\textbf{Pattern closedness checking.} A straightforward method for determining whether a frequent pattern $P$ is closed (or maximal) is to employ a postprocessing filtering technique. This technique starts by finding and storing in a set $S$ all frequent embedded patterns, and then performs a pairwise checking of patterns in $S$ to eliminate those patterns that are not closed (or maximal). Note that the problem of checking if one pattern is an embedded subpattern of another is NP-Complete \cite{KilpelainenM95}. This technique requires storing all the frequent patterns and doing O($|S|^2$) such checks. Clearly, this technique is not efficient since the number of closed and maximal patterns can be exponentially smaller than the number of  frequent patterns. 

An alternative way is to check each pattern $P$ for closedness as soon as it is discovered to be frequent. In this case one can enumerate all the possible embedded superpatterns having one more node than $P$, and then check whether there is a frequent superpattern $P'$ s.t. $L_{root}(P) = L_{root}(P|P')$. For maximality, we just need to check if the superpattern $P'$ is frequent. Although this method does not need to store any discovered patterns, it has to enumerate a large number of patterns, 
 and it is possible that a big number of superpatterns might be repeatedly generated and checked throughout the mining process.

Therefore, we develop a method which does not have the drawbacks of the techniques above.
During the frequent pattern mining process, this method identifies a subset of frequent patterns which is a superset of closed patterns, and are called locally closed patterns. A frequent pattern $P$ is {\em locally closed} if its root occurrence list is not equal to that of any pattern which can be obtained by joining $P$ (as a left or right operand) with a pattern from its class. A pattern $P$ is {\em locally maximal} if there are no canonical patterns in the class $[P]$. Obviously, a pattern which is not locally closed (or locally maximal) is not closed (or maximal). Consequently, we can find the ``globally'' closed patterns, by checking only locally closed patterns for closedness. Maximal patterns can be found similarly. This way, the search is narrowed to only those potentially closed (or maximal) patterns. Procedure $MineEmbPatterns$ of Fig. \ref{alg_TPMEmb} implements this method opportunistically.

\vspace*{.5ex}\noindent\textbf{Pruning non-closed patterns eagerly.} In order to prune non-closed patterns, we maintain a set ${\cal C}$ of all the patterns that are found to be closed w.r.t. frequent patterns discovered so far. Set ${\cal C}$ is initially empty. When all the possible expansions of a current pattern $P$ have been processed and $P$ has been found to be locally closed, it is compared with patterns in ${\cal C}$. It is added to ${\cal C}$, if there is no pattern $Q$  in {\cal C} which is a proper superpattern of $P$ and $L_{root}(P) = L_{root}(P|Q)$.  At the same time, patterns $Q$ in ${\cal C}$ that are sub-patterns of $P$ and satisfy the property $L_{root}(Q) = L_{root}(Q|P)$ are removed from ${\cal C}$. During the pattern comparison process, any pattern that is found to be a subpattern of another pattern is characterized as non-maximal. Procedure $CheckClosedMaxSubpattern$ of Fig. \ref{alg_TPMEmb} implements this process.

\begin{figure}[!t]
 \rule{\linewidth}{.5pt}

\emph{Input:} $minsup$ and set $\cal{L}$ of inverted lists of tree $T$.\\
\emph{Output:} set ${\cal C}$ of frequent closed and set ${\cal M}$ of frequent maximal embedded k-patterns in $T$ ($k>1$).\\

\vspace*{-2ex}
\begin{small}
 \algsetup{linenodelimiter=.}
  \begin{algorithmic}[1]

\STATE $F_1$ := \{frequent 1-patterns\};
\STATE $F_2$ := \{classes $[P]$ of frequent 2-patterns\};
\STATE ${\cal C}:=\emptyset$;
\FOR{(every $[P]\in F_2$)}
\STATE $MineEmbPatterns$($[P]$);
\ENDFOR
\STATE ${\cal M}$ := \{$P\mid P\in {\cal C} \;and \;P.isMax = true $\}
\end{algorithmic}

\vspace*{1ex}
\textbf{Procedure} $MineEmbPatterns$(Equivalence class $[P]$)\\
\vspace*{-3ex}
\begin{algorithmic}[1]
\FOR{(each $P_{x}^{i}\in [P]$)}
\IF{($P_{x}^{i}$ is in canonical form)}
\STATE $P_{x}^{i}.isLocallyClosed$ := true;
\STATE $[P_{x}^{i}]$ := $\emptyset$;
\FOR{(each $P_{y}^{j}\in [P]$)}
\FOR{(each expansion outcome $Q$ of $P_{x}^{i} \otimes P_{y}^{j}$)}

\STATE $OL(Q) := computeEmbOL(Q$, $P_{x}^{i}$, $P_{y}^{j}$, $OL(P_{x}^{i})$, $OL(P_{y}^{j})$);
\IF{($Q$ is frequent)}

\STATE add $Q$ to $[P_{x}^{i}]$;
\STATE let $L_{r_i}$,$L_{r_j}$, $L_{r_Q}$  be the emb. root occur. lists of $P_{x}^{i}$, $P_{y}^{j}$ and $Q$, respectively;
\IF{($L_{r_Q} = L_{r_i}$)}
\STATE $P_{x}^{i}.isLocallyClosed$ := false;
\ENDIF
\IF{($L_{r_Q} = L_{r_j}$)}
\STATE $P_{y}^{j}.isLocallyClosed$ := false;
\ENDIF
\ENDIF
\ENDFOR
\ENDFOR
\IF{($P_{x}^{i}.isLocallyClosed$ is true)}
\STATE $CheckClosedMaxSubpattern$($P_{x}^{i}$);
\ENDIF
\STATE $MineEmbPatterns$($[P_{x}^{i}]$)
\ENDIF
\ENDFOR
\end{algorithmic}

\vspace*{1ex}
\textbf{Procedure} $CheckClosedMaxSubpattern$($P$)\\
\vspace*{-3ex}
\begin{algorithmic}[1]

\FOR{(each $Q\in {\cal C}$)}
\IF{($Q\sqsubseteq_e P$)}
\STATE $Q.isMax$ := $false$;
\IF{($L(Q) = L(Q|P)$)}
\STATE remove $Q$ from ${\cal C}$;
\ENDIF
\ELSIF{($P\sqsubseteq_e Q$)}
\STATE $P.isMax$ := false;
\IF{($L(P) = L(P|Q)$)}
\STATE $P.isClosed$ := false;
\ENDIF
\ENDIF
\ENDFOR
\IF{($P.isClosed$)}
\STATE add $P$ to ${\cal C}$;
\ENDIF
\end{algorithmic}

\end{small}
\rule{\linewidth}{.5pt}
\caption{Algorithm {\em closedEmbTM-eager} for mining closed embedded tree patterns.}
\label{alg_TPMEmb}
\end{figure}

\subsection{The Closed Embedded Tree Pattern Mining Algorithm {\em closedEmbTM-eager}}
\label{subsection: closedEmbTM-eager}

We describe below the closed embedded tree pattern mining algorithm which checks for closedness eagerly. The algorithms is called {\em closedEmb\-TM-eager}  and is shown in Fig. \ref{alg_TPMEmb}. Algorithm {\em closedEmb\-TM-eager} first generates frequent 1-patterns (set $F_1$) and 2-patterns (set $F_2$). The equivalence class-based expansion technique (Section \ref{subsec:candGen}) is used to recursively constructs larger patterns by expanding each 2-pattern $\in F_2$ in a depth-first manner. While generating the patterns, {\em closedEmb\-TM-eager} uses the eager pattern closedness checking method described in Sec. \ref{subsection:Closeness} for reducing the pattern search space.

The outer procedure calls $Mine\-Emb\-Patterns$ on each frequent 2-pattern (lines 4-5).  $Mine\-Emb\-Patterns$ confirms that a pattern $P_{x}^{i}$ is in canonical form before expanding it (line 2). Subsequently, pattern $P_{x}^{i}$ is expanded through a join with each pattern $P_{y}^{j}\in [P]$ (if this is possible) and the occurrence list set of every join result is computed by evoking function $computeEmbOL$ (lines 5-7). Function $computeEmbOL$ computes the embedded occurrence lists for all the pattern nodes. It is an variation of the function $computeEmbOL$ introduced in \cite{WuT15,WuTSBDR18} for computing the pattern support extended so that it computes the embedded occurrence lists of all the nodes of the pattern and not only  the embedded occurrence list of the root of the pattern. Patterns that are discovered to be frequent are included in the class $[P_{x}^{i}]$ under construction (line 9).  Also, $P_{x}^{i}$ and $P_{y}^{j}$ can possibly be marked as non-locally closed (lines 12 and 14). After all patterns $P_{y}^{j}$ have been joined with $P_{x}^{i}$,  the class $[P_{x}^{i}]$ comprises all frequent patterns which share the common immediate prefix $P_{x}^{i}$. At this time, if $P_{x}^{i}$ has not been marked as non-locally closed, $Mine\-Emb\-Patterns$ calls procedure $CheckCloMaxSubpattern$ on $P_{x}^{i}$ to prune non-closed patterns (lines 15-16).  Finally, $Mine\-Emb\-Patterns$ is recursively called on $[P_{x}^{i}]$ to extract bigger frequent patterns which have a common prefix $P_{x}^{i}$ (line 17). This  process iterates recursively till all frequent patterns are produced.

%% file: texs/closedAlg-work.tex
\section{A Search Space Pruning Algorithm}
\label{sec:algoPrune}

Despite the early pattern closedness checking, Algorithm {\em closedEmbTM-eager} may generate numerous intermediate patterns which do not expand to become closed patterns. This motivates the need for developing an algorithm which exploits techniques for further pruning the generation of such non-promising intermediate patterns. We start by introducing these search space pruning techniques.

\subsection{Pattern Search Space Pruning Techniques}
\label{subsec:pruneRules}

\vspace*{.5ex}\noindent\textbf{Ordering the search space tree.} In order to introduce our search space pruning rules, we impose an order on the search space tree (Section \ref{subsec:candGen}).  We first define a linear order $\leq$ on tree pattern representations. Let there be a linear order $\leq$ on the labels of nodes in the input data tree.  Let $P_1$ and $P_2$ be two tree patterns representations whose roots are $r_1$ and $r_2$, respectively, and let $c_1^{r_1},\ldots,c_m^{r_1}$ and $c_1^{r_2},\ldots,c_n^{r_2}$ be, respectively, the lists of the children of $r_1$ and $r_2$. If $x$ is a node in a tree, $st(x)$ denotes the subtree (tree pattern representation) rooted at node $x$. Then, $P_1 \leq P_2$ iff:\\ 
$\bullet$ $lb(r_1) < lb(r_2)$, or\\
$\bullet$ $lb(r_1) = lb(r_2)$, and either: (a) $n \leq m$ and $st(c_i^{r_1}) = st(c_i^{r_2}),$ $\forall i \in [1, n]$, i.e., $P_2$ is equal to, or a prefix of $P_1$,
  or (b) $\exists j \in [1,min(m,n)]$ s.t. $st(c_i^{r_1})$ $= st(c_i^{r_2}),$ $\forall i,\; i < j$, and $st(c_j^{r_1})$ $< st(c_j^{r_2})$. \\

A pattern search tree becomes ordered when we enforce the linear order $\leq$ on sibling patterns of the tree. A pattern is created before its right sibling patterns in the ordered search space tree.

Our pruning rule is based on the observation that if it is known that a generated pattern and all (or some) of its descendant patterns in the pattern search tree are not closed, then these descendants need not be generated at all (they can be pruned from the pattern search tree). Our pruning rules are expressed using the concepts of child and cousin surrogates introduced below.

\vspace*{.5ex}\noindent\textbf{Child and cousin surrogates.} We start by defining child surrogates. We first define the concept of occurrence equivalent patterns. Let $P$ be an embedded subpattern of a pattern $P'$. If $e$ and $e'$ are embeddings of $P$ and $P'$ to a data tree $T$, respectively, $e'$ is an {\em extension} of $e$ if there exists  an embedding $m$ of $P$ into $P'$ s.t. $e(X)=e'(m(X))$, for every node $X$ in $P$.

\begin{definition}[Pattern occurrence equivalence]
\label{def:occ.equivalence}
Let $P$ be an embedded subpattern of $P'$, and let $M(P, P')$ represent the set of possible embeddings from $P$ to $P'$. Patterns $P$ and $P'$ are {\em occurrence equivalent} iff for every embedded occurrence $e$ of $P$ to $T$ there exists an embedded occurrence of $P'$ to $T$ which is an extension to $e$ for some $m \in M(P, P')$. We use the notation $P \equiv P'$ for occurrence equivalent patterns $P$ and $P'$.
\label{def:close}
\end{definition}

The concept of occurrence equivalence was first introduced in \cite{YanH03} for mining induced subgraphs, and was also later used in \cite{ChiXYM05}, with a different name, for mining induced subtrees. Our definition generalizes these definitions, as it considers patterns related through embeddings not merely isomorphisms.

Since non-canonical patterns are not further expanded, we focus on finding canonical patterns. We call an expansion ``canonical'' if the expansion outcome is a canonical pattern. In the following discussion, except when stating differently, we assume canonical expansions.

\begin{figure*}[!t]
    \centering%
     \scalebox{0.8}{ \epsfig{file=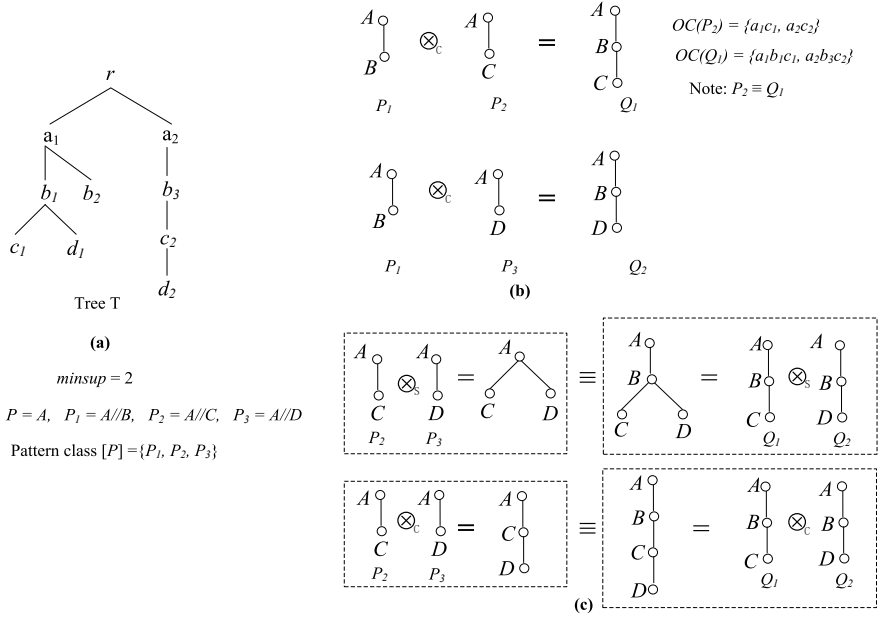}}%
    \caption{An example of child surrogate: $P_1$ is a child surrogate of both $P_2$.}
 \label{fig:ruleExample1}
\end{figure*}

Consider the example of Fig. \ref{fig:ruleExample1}. In this figure, focus on the data tree $T$ and the frequent patterns (assuming minimum support 2) $P_0\,=\,A, P_1=A//B, P_2 = A//C$, and $P_3 = A//D$ shown in Fig. \ref{fig:ruleExample1}(a). One can see that the equivalence class $[A]$ of $A$ is $ \{A//B, A//C, A//D \}$. Let's now look at a property of patterns $P_1 = A//B$ and $P_2 = A//C$ in $[A]$. Observe that in every embedding of pattern $P_2 = A//C$ to the $T$, $B$ is an ancestor of $C$ below node $A$ in the data tree. In other words, every time $P_2$ has an embedding to $T$, pattern $Q_1 = P_1\otimes_c P_2=A//B//C$ also has an embedding to $T$ which coincides with $P_2$ on their common part (nodes $A$ and $C$).  This means that $P_2 \equiv Q_1$. Fig. \ref{fig:ruleExample1}(b) shows the occurrence sets $OC(P_2)$ and $OC(Q_1$) of $P_1$ and $Q_1,$ respectively.

Further, let's see  another property of patterns $P_1$ and $P_2$ in $[A]$. Consider pattern $P_3 = A//D \in [A]$ shown in Fig. \ref{fig:ruleExample1}(c). It is, $P_2 < P_3$. Observe that in every embedding of the pattern $P_2 \otimes_c P_3$ (or $P_2 \otimes_s P_3$) to $T$, node $B$ is an ancestor of node $C$ below the common part of $P_2 = A//C$ and $P_3 = A//D$ (which is node $A$) in the data tree. In other words, every time $P_2 \otimes_c P_3$ (resp. $P_2 \otimes_s P_3$) has an embedding to $T$, pattern $Q_1 \otimes_c Q_2$ (resp. $Q_2 \otimes_s Q_3$) also has an embedding to $T$ which coincides with  $P_2 \otimes_c P_3$ (resp. $P_2 \otimes_s P_3$)  on their common part (nodes $A$, $C$ and $D$). That is, $(P_2 \otimes_c P_3) \equiv (Q_1 \otimes_c Q_2)$  (resp. $(P_2 \otimes_c P_3) \equiv (Q_1 \otimes_c Q_2)$ ). When this property holds for patterns $P_1$ and $P_2$ not only in relation to pattern $P_3$ but also  in relation to every pattern $P_z$ ($P_2 \leq P_z$) in the class $[A]$ in which $P_1$ and $P_2$ belong, we say that $P_1 = A//B$ is a surrogate pattern of $P_2 = A//C$. A formal definition follows.

\begin{definition}
\label{def:csurrogate}
Consider two distinct patterns $P_{x}^i$ and $P_{y}^i$ in the class $[P]$, such that $P_{x}^i < P_{y}^i$, and let $Q$ be the child expansion of $P_x$ with $P_y$ ($Q = P_x^i \otimes_c P_y^i$). We say that $P_{x}^i$ is a {\em child surrogate} of $P_{y}^i$, if:

\begin{itemize}
\item [(a)] $P_{y}^i \equiv Q$, and
\item [(b)] for any $P_{z}^i\in[P]$ such that $P_{y}^ i\leq P_{z}^i$, $(P_y^i \otimes_s P_z^i) \equiv (P_x^i \otimes_c P_y^i) \otimes_s (P_x^i \otimes_c P_z^i)$ or $(P_y^i \otimes_s P_z^i) \equiv (P_x^i \otimes_c P_y^i) \otimes_s (P_x^i \otimes_s P_z^i)$.
\end{itemize}
\end{definition}

\begin{figure*}[!t]
    \centering%
     \scalebox{0.8}{ \epsfig{file=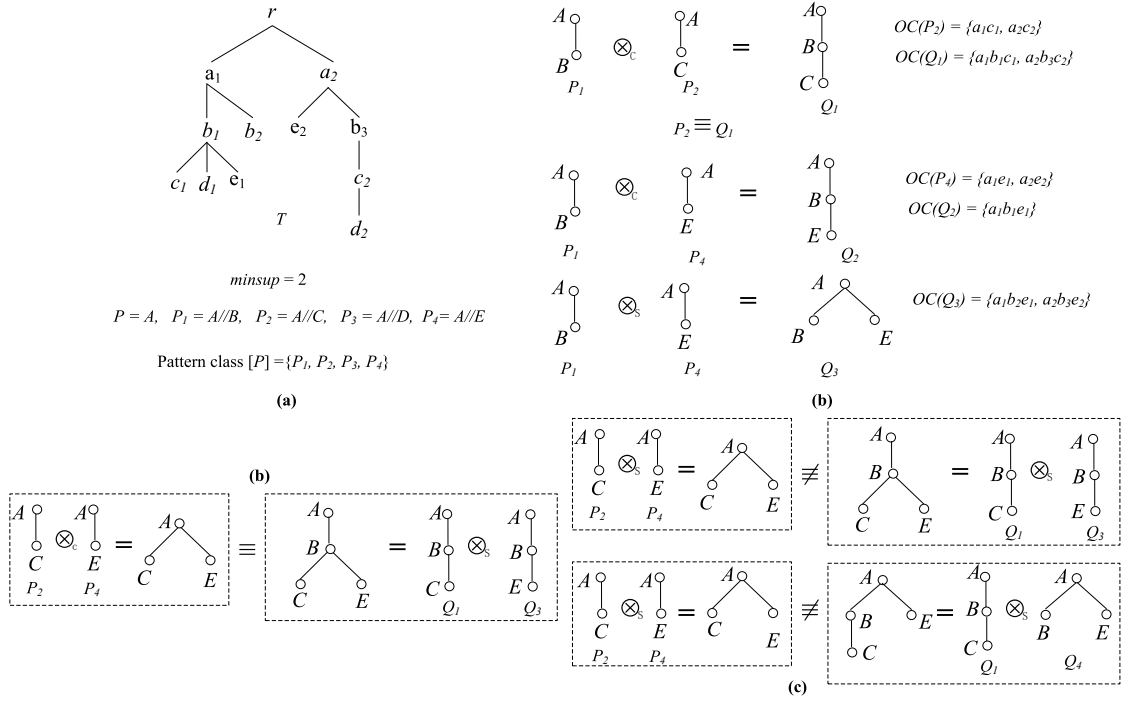}}%

       \caption{A non-example of child surrogate: $P_1$ is not a child surrogate of $P_2$.}
 \label{fig:ruleExample2}
\end{figure*}

\begin{example}
\label{ex:csurrogate}
As mentioned above, Fig. \ref{fig:ruleExample1} shows an example of a child surrogate: $P_1$ is a child surrogate of $P_2$. Fig. \ref{fig:ruleExample2} shows a non-example of a child surrogate: $P_1$ is not a child surrogate of $P_2$. Indeed, as shown in the figure even though the condition (a) of Definition \ref{def:csurrogate} is satisfied, there is pattern $P_4 = A//E$ in the class $[A]$ such that $P_2 > P_4$, $(P_2 \otimes_s P_4) \not\equiv (P_1 \otimes_c P_2) \otimes_s (P_1 \otimes_c P_4)$ and $(P_1 \otimes_s P_4) \not\equiv (P_1 \otimes_c P_2) \otimes_s (P_1 \otimes_s P_4)$. That is, condition (b) of Definition \ref{def:csurrogate} is violated.
\end{example}

If  $P_x^i$ is a child surrogate of $P_y^i$ then, clearly, $P_y^i$ is not closed and the same is true for any canonical expansion (child or cousin) of $P_y^i$. We will show later that every node in the pattern tree rooted at $P_y^i$ is also not closed and therefore, these nodes need not be generated.

Analogously to child surrogates, we define cousin surrogates.

\begin {definition}
\label{def:cnsurrogate}
Consider two distinct patterns $P_{x}^i$ and $P_{y}^j, \; i \geq j,$ in the class $[P]$, such that $P_{x}^i < P_{y}^j$, and let $Q$ be the cousin expansion outcome of $P_x^i$ with $P_{y}^j$ ($Q = P_{x}^{i}\otimes_s P_{y}^j$). We say that $P_{x}^i$ is a {\em cousin surrogate} of $P_{y}^j$, if:
\vspace*{-1ex}
\begin{itemize}
\item [(a)] $P_{y}^j \equiv Q$, and
\item [(b)] for any $P_{z}^k\in[P],$ such that $P_{y}^j\leq P_{z}^k$, $P_y^j \otimes_s P_z^k  \equiv  (P_x^i \otimes_s P_y^j) \otimes_s (P_x^i \otimes_s P_z^k).$
\end{itemize}
\end{definition}

\begin{figure*}[!t]
    \centering%
     \scalebox{0.8}{ \epsfig{file=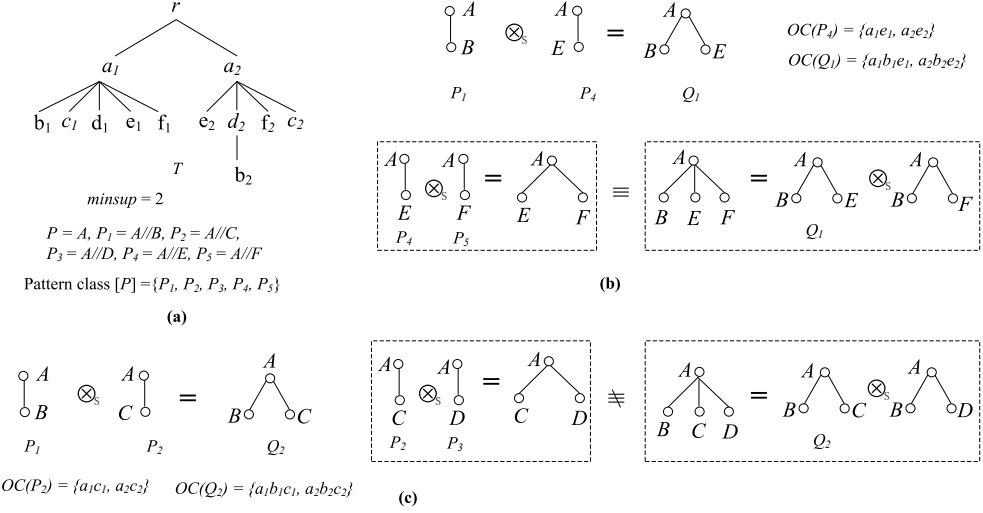}}%
     \caption{Examples and non-examples of cousin surrogates: $P_1$ is not a cousin surrogate of $P_2$, but it is a cousin surrogate of $P_4.$}
 \label{fig:ruleExample3}
\end{figure*}

\begin{example}
Consider the data tree $T$ of Fig. \ref{fig:ruleExample3}(a) and the frequent patterns (assuming minimum support 2) $P=A, P_1=A//B, P_2 = A//C, P_3 = A//D, P_4 = A//E$, and $P_5 = A//F$. The equivalence class $[A]$ of $A$  is $ \{P_1, P_2, P_3, P_4, P_5 \}$.  Pattern $P_1$ is a cousin surrogate of $P_4$ since, as shown in Fig.~\ref{fig:ruleExample3}(b), both conditions of Definition \ref{def:cnsurrogate} are satisfied: (1)~$P_4 \equiv Q_2\; (= P_1 \otimes P_4)$, and (2) for $P_5 \in[A]$ ($P_5$ is the only pattern in the class $[A]$ which follows $P_4$),
$P_4 \otimes_s P_5  \equiv  (P_1 \otimes_s P_4) \otimes_s (P_1 \otimes_s P_5).$ Note that $P_4 \otimes_c P_5$ does not have any occurrences in the data tree $T$.

In contrast, pattern $P_1$ is not a cousin surrogate of $P_2$. Indeed, as shown in Fig.~\ref{fig:ruleExample3}(c), even though $P_2 \equiv Q_3\;  (= P_1 \otimes P_2$), for the pattern $P_3$ in the class $[A]$ ($P_1 \prec P_3$), $P_2 \otimes_s P_3  \not\equiv  (P_1 \otimes_s P_2) \otimes_s (P_1 \otimes_s P_3).$ That is, condition (2) of Definition \ref{def:cnsurrogate} is not satisfied.

\end{example}

Under certain circumstances, checking for cousin surrogates can be simplified as the next proposition shows.

\begin{proposition}
\label{prop:cnsurrogate}
Consider two distinct patterns $P_{x}^i$ and $P_{y}^j, \; i \geq j,$ in the class $[P]$, such that $P_{x}^i < P_{y}^j$. If $P_{y}^j \equiv P_x^i \otimes_s P_y^j$, and for any $P_{z}^k\in[P]$ such that $P_{y}^j < P_{z}^k$, we have $i>k$, then  $P_{x}^i$ is a cousin surrogate of $P_{y}^j$.
\end{proposition}

\begin{figure*}[!t]
    \centering%
     \scalebox{0.8}{ \epsfig{file=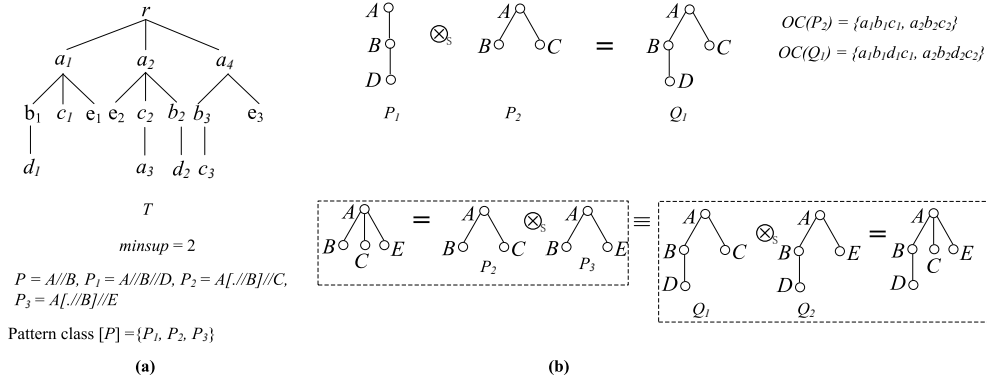}}%
     \caption{The rightmost leaf position of $P_1$ is larger than that of $P_3:$ $P_1$ is a cousin surrogate of $P_2.$}
 \label{fig:ruleExample4}
\end{figure*}

\begin{example}
Consider the data tree $T$ of Fig.  \ref{fig:ruleExample4}(a) and the frequent patterns (assuming minimum support 2) $P, P_1, P_2,$ and $ P_3$. The equivalence class $[P]$ of $P=A//B$  is $ \{P_1, P_2, P_3 \}$. Patterns $P_1$ and $P_2$ satisfy condition (a) of Definition \ref{def:cnsurrogate} as  $P_2 \equiv  P_1 \otimes_s P_2$. There is only one pattern (pattern $P_3$) in $[P]$ which follows pattern $P_1$ ($P_1 < P_3$). Observe that the position of the rightmost leaf of $P_3$ (position 0) is inferior to the position of the rightmost leaf of $P_1$ (position 1). In this case, we can conclude that condition 2 of definition  \ref{def:cnsurrogate} is satified without checking (and that $P_1$ is a cousin surrogate of $P_2$). Indeed, in every occurrence of $P_2 \otimes_s P_3$, $D$ occurs under $B$ in $T$ ($P_2 \otimes_c P_3$ does not have any occurrence in $T$) .
\end{example}

\vspace*{.5ex}\noindent\textbf{Pattern search tree pruning.} We exploit the concepts of child and cousin surrogates to design a technique which allows pruning branches (subtrees) of patterns from the pattern search tree. We start by providing a lemma.

\begin{lemma}
\label{lem:rule1}
If a pattern has a child or a cousin surrogate, then neither itself nor anyone of its (child or cousin) expansion can be closed.
\end{lemma}

Based on lemma \ref{lem:rule1}, we design a pruning rule:

\vspace*{1ex}\noindent{\bf Prunning Rule:} If a pattern has a child or cousin surrogate, then none of its descendants in the pattern search tree needs to be generated.

\vspace*{1ex}The correctness of the prunning rule is shown by the next proposition.

\begin{proposition}
\label{prop:rule1}
If a pattern has a child or a cousin surrogate, then neither itself nor anyone of its descendants in the pattern search tree can be closed (or maximal).
\end{proposition}

For instance, in the example of Fig. \ref{fig:ruleExample1}, pattern $P_1$ is shown to be a child surrogate of pattern $P_2$. Therefore, $P_2$ does not need to be expanded: none of its descendants will be closed (or maximal). The same holds for pattern $P_4$ in the example of Fig. \ref{fig:ruleExample3} since it was shown there that pattern $P_1$ is a cousin surrogate of $P_4$.

\subsection{\mbox{The Closed and Maximal Embedded Pattern Mining Algorithm  {\em closedEmbTM\--prune}}}

We now present {\em closedEmbTM-prune}, an efficient algorithm which allows extracting all  closed and maximal embedded frequent patterns. The pseudo-code for\linebreak {\em closedEmbTM-prune} for  patterns with at least two nodes appears in Fig. \ref{alg_TPMEmbCLMX}. The main computation is performed by Procedure {\em MineClosedMaxEmbPatterns} called on frequent \mbox{2-patterns} (Line 3-4). While the high level structure of {\em MineClosedMaxEmbPatterns} is similar to Procedure $Mine\-Emb\-Patterns$ 
(Fig. \ref{alg_TPMEmb}), {\em MineClosedMaxEmbPatterns} comprises several important extensions which are described below.

\vspace*{.5ex}\noindent\textbf{Class element ordering.} As mentioned in Section \ref{subsec:pruneRules}, in order to facilitate pattern pruning, we impose a tree order among sibling nodes in the pattern search tree. To implement the ordering efficiently, we maintain the elements $P_x^i$ in every class primarily ordered on the position $i$ (in descending order) and secondarily by the node label $x$ (in ascending order). It is not difficult to see that this class element ordering coincides with the tree order defined in Section \ref{subsec:candGen}. Given the sorted element list of a class, for each pair of elements, the equivalence class expansion process first tries to apply child expansion (Lines 7-8), and then cousin expansion (Lines 10-11). The class under expansion is associated with two lists for storing the outcomes of child and cousin expansions. The expansion outcomes are added into their corresponding lists in the order they are generated. After all the expansion operations for the current class are finished, the cousin expansion list is appended to the child expansion list. This expansion process guarantees that the candidate generation phase returns an ordered set of patterns (no explicit ordering is needed).

\begin{figure}[!t]
 \rule{\linewidth}{.5pt}

\emph{Input:}  frequency threshold $minsup$ and inverted lists $\cal{L}$ of tree $T$.\\
\emph{Output:}  set ${\cal C}$ of the closed embedded patterns,   set ${\cal M}$ of the maximal embedded patterns in $T$ which have at least two nodes.\\
\vspace*{-2ex}
\begin{small}
 \algsetup{linenodelimiter=.}
  \begin{algorithmic}[1]

\STATE $F_2$ := \{classes $[P]$ of frequent 2-patterns\};
\STATE ${\cal C}:=\emptyset$;
\FOR{(each $[P]\in F_2$)}
\STATE $MineClosedMaxEmbPatterns$($[P]$, ${\cal C}$);
\ENDFOR
\STATE ${\cal M}$ := \{$P\mid P\in {\cal C} \;and \;P.isMax = true $\}
\RETURN ${\cal C}$, ${\cal M}$;
\end{algorithmic}

\vspace*{1ex}
\textbf{Procedure} $MineClosedMaxEmbPatterns$(Pattern equivalence class $[P]$, pattern set ${\cal C}$)\\
\vspace*{-3ex}
\begin{algorithmic}[1]
\FOR{(each $P_x^i\in [P]$ in ascending order)}

\IF{($P_{x}^{i}$ is non-canonical or a cousin surrogate of $P_x^i$ has been identified using Prop. \ref{prop:cnsurrogate})}
\STATE continue;
\ENDIF
\STATE $[P_{x}^{i}]$ := $\emptyset$;

\FOR{(each $P_{z}^{j}\in [P]$ in ascending order)}
\STATE $ComputeExpansion$($P_{x}^{i}$, $P_{z}^{j}$, $P_{x}^{i} \otimes_c P_{z}^{j}$, $[P_{x}^{i}]$);

\STATE $ComputeExpansion$($P_{x}^{i}$, $P_{z}^{j}$, $P_{x}^{i} \otimes_s P_{z}^{j}$, $[P_{x}^{i}]$);
\IF{($i=j$ and $P_{x}^i \leq P_{z}^{j}$)}
\FOR{(every $P_{y}^{j}\in[P]$ s.t. $P_{x}^i < P_{y}^{j} \leq P_{z}^{j}$)}
\IF{($P_{x}^{i}$ is in the child candidate list of $P_{y}^{j}$, and $P_{z}^{j}$ violates the conditions of Prop. \ref{prop:csurrogate})}
\STATE remove $P_{x}^{i}$ from the child candidate list of $P_{y}^{j}$;
\ENDIF
\IF{($P_{x}^{i}$ is in the cousin candidate list of $P_{y}^{j}$, and $P_{z}^{j}$ violates the conditions of Prop. \ref{prop:cnurrogate2})}
\STATE remove $P_{x}^{i}$ from the cousin candidate list of $P_{y}^{j}$;
\ENDIF

\ENDFOR
\ENDIF

\ENDFOR

\IF{($P_{x}^{i}.isLocallyClosed$ is $true$)}
\STATE $CheckClosedMaxSubpattern$(${\cal C}$, $P$);
\ENDIF

\IF{(no pattern in the candidate child and cousin surrogate lists of $P_x^i$ is a surrogate of  $P_x^i$)}
\STATE $MineClosedMaxEmbPatterns$($[P_{x}^{i}]$, ${\cal C}$)
\ENDIF
\ENDFOR
\end{algorithmic}

 \vspace*{1ex}
\textbf{Procedure} $ComputeExpansion$($P_{x}^{i}$, $P_{y}^{j}$, $Q$, $[P_{x}^{i}]$)\\
\vspace*{-3ex}
\begin{algorithmic}[1]
\STATE $OL(Q)$ := $computeEmbOL$($Q$, $P_{x}^{i}$, $P_{y}^{j}$, $OL(P_{x}^{i})$, $OL(P_{y}^{j})$);
\IF{($OL(Q)$ is frequent)}
\STATE add $Q$ to $[P_{x}^{i}]$;
\IF{($Q$ is in canonical form)}
\STATE $FindSurrogateCandidate$($P_{x}^{i}$, $P_{y}^{j}$, $Q$);
\ENDIF
\ENDIF

\end{algorithmic}

\end{small}
\rule{\linewidth}{.5pt}
\caption{Closed and maximal embedded pattern mining Algorithm {\em closedEmpTM-prune}.}
\label{alg_TPMEmbCLMX}
\vspace*{-5ex}
\end{figure}

\vspace*{.5ex}\noindent\textbf{Pattern expansion pruning.} The pruning rule presented above is applied throughout the pattern mining process.  The pruning rule requires identifying child/cousin surrogates of patterns during the mining process. In order to determine whether a given pattern $P$ has a child or cousin surrogate, we  first need to identify a pattern preceding\footnote{We say that a pattern $Q$ {\em precedes} a pattern $P$, (and that $P$ {\em follows} $Q$) in the ordered search space tree if $Q \le P$.} $P$ which satisfies condition (a) in the definition \ref{def:csurrogate} or \ref{def:cnsurrogate}. Then we expand $P$ with each of the patterns that follow $P$ in the class ordering, in order to check if condition (b) of definitions \ref{def:csurrogate} and \ref{def:cnsurrogate} is satisfied.  Proposition \ref{prop:cnsurrogate}  allows identifying a cousin surrogate of $P$ without exhaustively expanding $P.$

\vspace*{1ex}\noindent\textbf{Improving the pruning process.} Next, we introduce two propositions which can help  avoiding part of the computation while determining child and cousin surrogates. Let  $\Pi_{P_{x}^i}(OC(P_x^i \otimes_c P_y^i))$ denote the projection of the embedded occurrence set $OC(P_x^i \otimes_c P_y^i)$ on the attributes corresponding to the nodes of $P_x^i$.

\begin{proposition}
\label{prop:csurrogate}
If $P_{x}^i$ is a child surrogate of $P_{y}^i$ then for any $P_{z}^i\in[P]$ such that $P_{y}^i<P_{z}^i$ and $\Pi_{P}(P_{y}^i) = \Pi_{P}(P_z^i)$, $\Pi_{P_{x}^i}(OC(P_x^i \otimes_c P_y^i))\subseteq \Pi_{P_{x}^i}(OC(P_{x}^i \otimes_c P_{z}^i))$ or $\Pi_{P_x^i}(OC(P_x^i \otimes_c P_y^i))\subseteq \Pi_{P_x^i}(OC(P_x^i \otimes_s P_z^i))$.
\end{proposition}

Going back to the example of Figure \ref{fig:ruleExample2}, we can see that the conditions of Proposition \ref{prop:csurrogate} are not satisfied when $P_{x}^i = P_1$,  $P_{y}^i = P_2$ and $P_z^i = P_4$. Therefore, we can conclude that $P_1$ is not a child surrogate of $P_2$ without recurring to exhaustively checking the conditions of Definition \ref{def:csurrogate}.

\begin{proposition}
\label{prop:cnurrogate2}
 If $P_{x}^i$ is a cousin surrogate of $P_{y}^i$, then for any $P_{z}^{i}\in[P]$ such that $P_{y}^{i} < P_{z}^{i}$, $\Pi_{P_{x}^{i}}(OC(P_x^i \otimes_s P_y^i))$ is disjoint from both $\Pi_{P_{x}^{i}}(OC(P_{x}^{i} \otimes_c P_{z}^{i}))$ and $\Pi_{P_{x}^{i}}(OC(P_{z}^{i} \otimes_c P_{x}^{i})).$
\end{proposition}

In the example of Fig. \ref{fig:ruleExample3},  the conditions of Proposition \ref{prop:cnurrogate2} are not satisfied when $P_{x}^i = P_1$,  $P_{y}^i = P_2$ and $P_z^i = P_3$. Therefore, we can conclude that $P_1$ is not a child surrogate of $P_2$ avoiding the exhaustive checking of the conditions of Definition \ref{def:cnsurrogate}.

\vspace*{1ex}\noindent\textbf{Algorithm description.}  We describe now in more detail how the pruning rule is integrated into the mining process. Let $P_{x}^{i}$ be the pattern under consideration in class $[P]$. Pattern $P_{x}^{i}$ is not expanded if it is non-canonical or a cousin surrogate of it has been identified earlier using Proposition \ref{prop:cnsurrogate} (Lines 2-3 of {\em MineClosedMaxEmbPatterns}). Otherwise, Procedure {\em ComputeExpansion} is invoked to compute the joins $P_{x}^{i} \otimes P_z^k$ of $P_x^i$ with a patern $P_z^k$ in $[P]$ (Lines 6-7). Procedure {\em ComputeExpansion}, in turn, calls {\em computeEmbOL} which computes the embedded occurrence lists of the join outcome. Let $Q$ denote the outcome of $P_{x}^{i} \otimes P_{z}^{j}$.  If $Q$ is frequent and canonical, {\em CompExpansion} goes on to check, via a call to {\em FindSurrogateCandidate}, shown in Figure \ref{alg_findSurrogates},  whether: (1) $P_{x}^{i}$ and $P_{z}^{j}$ are locally closed (Lines 2-5), and (2) $P_{x}^{i}$ and  $P_{z}^{j}$ satisfy Condition (a) of the child/cousin surrogate definition (Line 6). To facilitate this checking, each pattern in an equivalence class is associated with two lists recording its child and cousin surrogate candidates  (Lines 8 and 12).   Pattern $P_{x}^{i}$ can be determined to be a cousin surrogate when the condition of Proposition \ref{prop:cnsurrogate} is satisfied (Lines 9-10). A child/cousin surrogate candidate added by {\em FindSurrogateCandidate} to a candidate list may be removed later, when a pattern in the same class disqualifies it (Lines 8-13 of {\em MineClosedMaxEmbPatterns}). The patterns in the class $[P_{x}^{i}]$ are expanded only when no candidate in the child and cousin surrogate candidate lists of $P_{x}^{i}$ is indeed a surrogate of  $P_{x}^{i}$ (Lines 16-17 of Procedure $MineClosedMaxEmbPatterns$).

\begin{figure}[!t]
 \rule{\linewidth}{.5pt}
\begin{small}
 \algsetup{linenodelimiter=.}

\textbf{Procedure} $FindSurrogateCandidate$($P_{x}^{i}$, $P_{y}^{j}$, $Q$)\\
\vspace*{-3ex}
\begin{algorithmic}[1]

\STATE let $L_{r_i}$,$L_{r_j}$, $L_{r_Q}$  be the embedded root occurrence lists of $P_{x}^{i}$, $P_{y}^{j}$ and $Q$, respectively;
\IF{($L_{r_Q} = L_{r_i}$)}
\STATE $P_{x}^{i}.isLocallyClosed$ := false;
\ENDIF
\IF{($L_{r_Q} = L_{r_j}$)}
\STATE $P_{y}^{j}.isLocallyClosed$ := false;
\ENDIF

\IF{($P_{y}^{j} \equiv Q$)}
\IF{($Q = P_{x}^{i} \otimes_s P_{y}^{j}$)}
\STATE add $P_{x}^{i}$ to the cousin surrogate candidate list of $P_{y}^{j}$;
\IF{($i>j$)}
\STATE mark $P_{x}^{i}$ as a cousin surrogate of $P_{y}^{j}$;
\ENDIF
\ELSIF{($Q = P_{x}^{i} \otimes_s P_{y}^{j}$)}
\STATE add $P_{x}^{i}$ to the child surrogate candidate list of $P_{y}^{j}$;
\ENDIF
\ENDIF

\end{algorithmic}

\end{small}
\rule{\linewidth}{.5pt}
\caption{Find candidate child and cousin surrogates.}
\label{alg_findSurrogates}
\end{figure}

\begin{theorem}
\label{theo:correctnessEmbCLMX}
Algorithm {\em closedEmpTM-prune} produces all and only closed and maximal embedded tree patterns.
\end{theorem}

\vspace{1.5ex}\noindent\textbf{Proof.}  Algorithm {\em closedEmpTM-prune} correctly identifies all the closed frequent embedded tree patterns since its search is based on a complete traversal of the pattern search tree. The only branches that are pruned are those that either do not have sufficient support or those that have a child or cousin surrogate (Proposition \ref{prop:rule1} establishes the correctness of the latter pruning). In addition, throughout the mining process, {\em closedEmpTM-prune} maintains a set ${\cal C}$ that stores all the locally closed patterns that are found to be closed w.r.t. frequent patterns discovered so far. A locally closed pattern $P$ is inserted into ${\cal C}$ only if Procedure {\em CheckClosedMaxSubpattern} cannot find any embedded superpattern in ${\cal C}$ which violates the closedness of $P$. If $P$ is inserted into ${\cal C}$, all embedded subpatterns of $P$ which Procedure {\em CheckClosedMaxSubpattern} identifies as non-closed are eliminated from ${\cal C}$. The closedness checking process also identifies and marks non-maximal patterns in ${\cal C}$. When the mining process terminates, ${\cal C}$ contains exactly all the closed patterns, while the non-marked patterns in ${\cal C}$ are the maximal patterns. \hfill$\Box$

%% file: texs/experiments.tex
\section{Experimental Evaluation}
\label{sec:experiments}

\noindent\textbf{Algorithms in comparison.} To evaluate our approach, we have designed different experiments. We implemented and compared our algorithms {\em closedEmbTM-eager} and {\em closedEmbTM-prune}. To the best of our knowledge, there is no previous algorithm for mining closed and maximal embedded patterns from data trees.  Therefore, we compared the performance of our two algorithms with a baseline algorithm which first computes the embedded tree patterns and then filters out non-closed (or non-maximal) ones. We call this post-processing algorithm {\em closedEmb\-TM-base}. In order to compute frequent embedded patterns, {\em closedEmb\-TM-base} extends the algorithm {\em embTM} \cite{WuT15,WuTSBDR18} which mines frequent embedded patterns from data trees so that it computes the occurrence lists not only of the root but also of the other nodes of a pattern. The pattern node occurrence lists are used in the post-processing phase for checking pattern closedness.

In order to contrast the performance of mining closed embedded patterns with mining frequent embedded pattrens, we also compare these three algorithms with {\em sleuth} \cite{Zaki05}, which is a state-of-the-art unordered embedded tree mining algorithm. The goal of this algorithm was to mine embedded patterns from small tree collections. In order to make the comparison meaningful, {\em sleuth} was slightly altered so that it mines patterns from a single large tree. This was achieved  by having {\em sleuth} adopting {\em root frequency} instead of {\em document frequency} as the support of a pattern. The performance of {\em sleuth} is not affected by this adaptation. In fact, by design, all the embedded occurrences of a pattern under consideration in a data tree are computed by {\em sleuth}. Note that algorithm {\em embTM} was shown to outperform Sleuth \cite{WuT15,WuTSBDR18} but we included also Sleuth in this comparison since  {\em embTM} is employed in {\em closedEmb\-TM-base}.

\subsection{Experimental Setting}
\label{subsec:settings}

We implemented our approach in Java. For the experiments we ran JVM 1.8 on a workstation having an Intel Xeon CPU 3565 @3.20 GHz processor. 

\vspace*{.5ex}\noindent\textbf{Datasets.}  For the experiments we employed three datasets with diverse structural features which are popular in pattern mining. The Table \ref{tab:data_stat} below summarizes teir main characteristics.

\begin{table}[!h]
       \center
      \begin{tabular}[b]{|c|c|c|c|c|}
            \hline
                    Dataset&Max/Avg depth		&Tot. \#nodes&\#labels		&\#paths\\
            \hline
             \hline
            XMark	&13/6.4&180769	&24245	& 138840\\	
              \hline
                DBLP	&7/3.6	&7146542 &941664 &6410975\\
            \hline
            CSlogs	&86/4.4	&772188	&13355	&59691 (\#trees)\\
            \hline

        \end{tabular}
   \caption{Dataset statistics.}
    \label{tab:data_stat}
  \vspace*{-3ex}
\end{table}

The {{\em XMark}}\footnote{\scriptsize{http://xml-benchmark.org}} dataset models an auction website and has been used as a benchmark. The dataset has many regularly structured patterns and is deep. It also comprises few recursive nodes (i.e., nodes which coexist on the same path with other nodes having the same label).

The {{\em DBLP}}\footnote{\scriptsize{http://dblp.uni-trier.de/xml/}} dataset provides bibliographic information and is a real dataset. The dataset is bushy, flat and shallow. It comprises few regularly structured patterns.  As XMark,  it includes very few recursive elements.

The {{\em $CSlogs$}}\footnote{\scriptsize{http://www.cs.rpi.edu/$\sim$zaki/software/}} dataset is presented in \cite{Zaki05tkde} and comprises real data. It contains navigation trees of different users on the website of the RPI CS department. The dataset involves 13355 distinct web pages which are accessed in 59,691 trees (with an average tree size of 12.94).

\begin{figure*}[!t]
       \center
       \subfigure[Number of frequent embedded patterns vs. sizes]{ \scalebox{0.45}{ \label{fig:xmpatlen} \epsfig{file=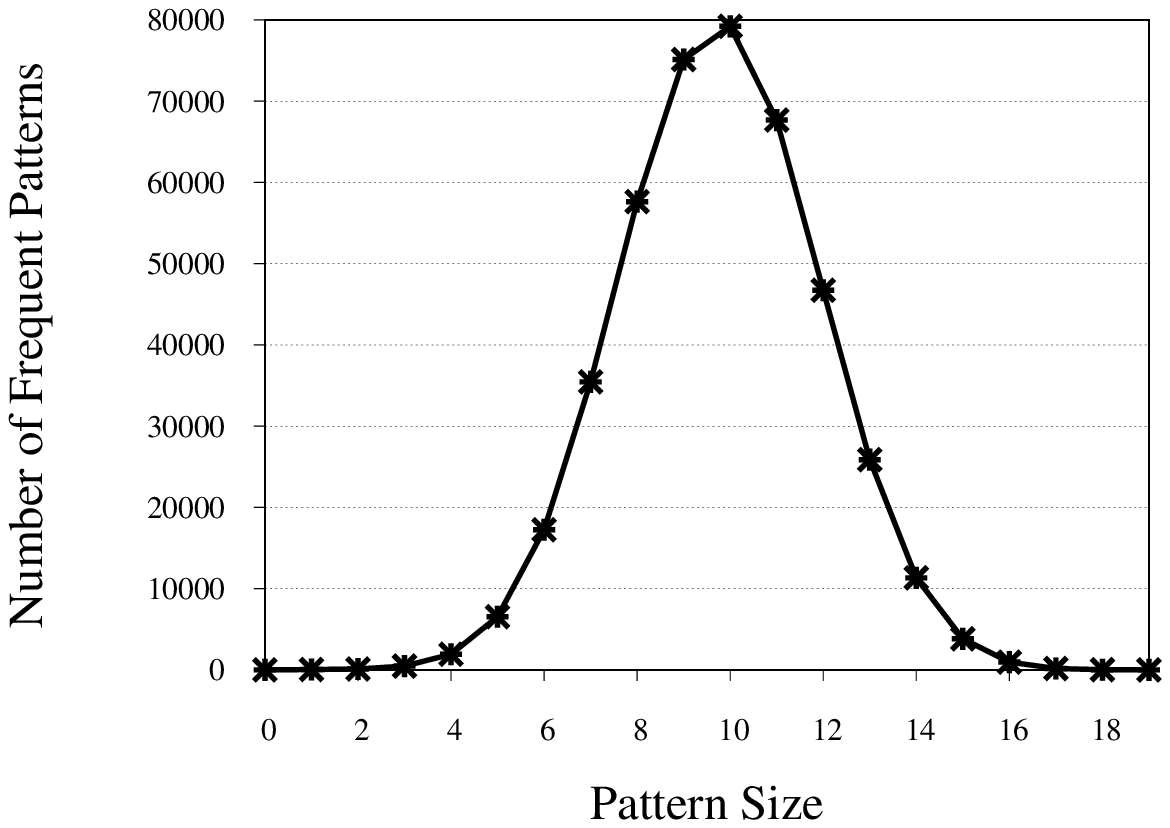} } }
        \subfigure[Number of closed and max patterns vs. sizes]{ \scalebox{0.45}{ \label{fig:xmpatlen2} \epsfig{file=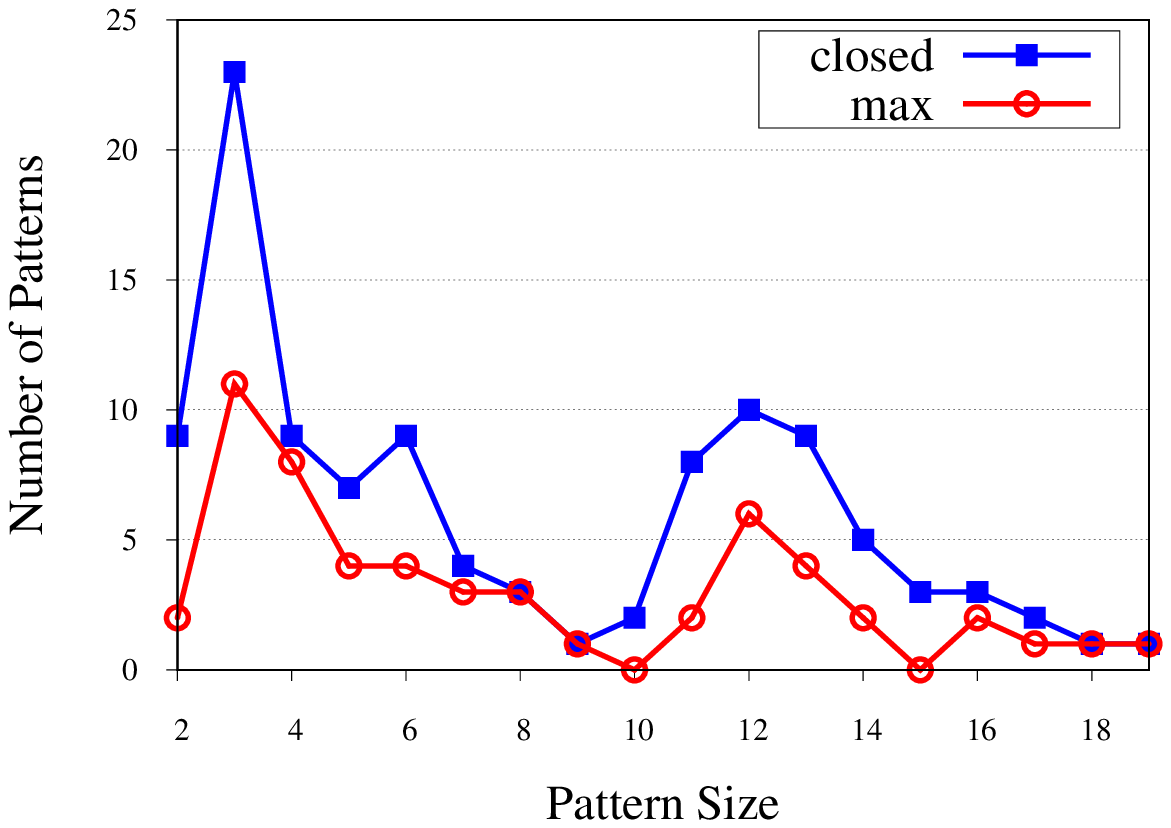}}}
        \subfigure[Number of patterns vs. minsup]{ \scalebox{0.45}{ \label{fig:xmpatsup} \epsfig{file=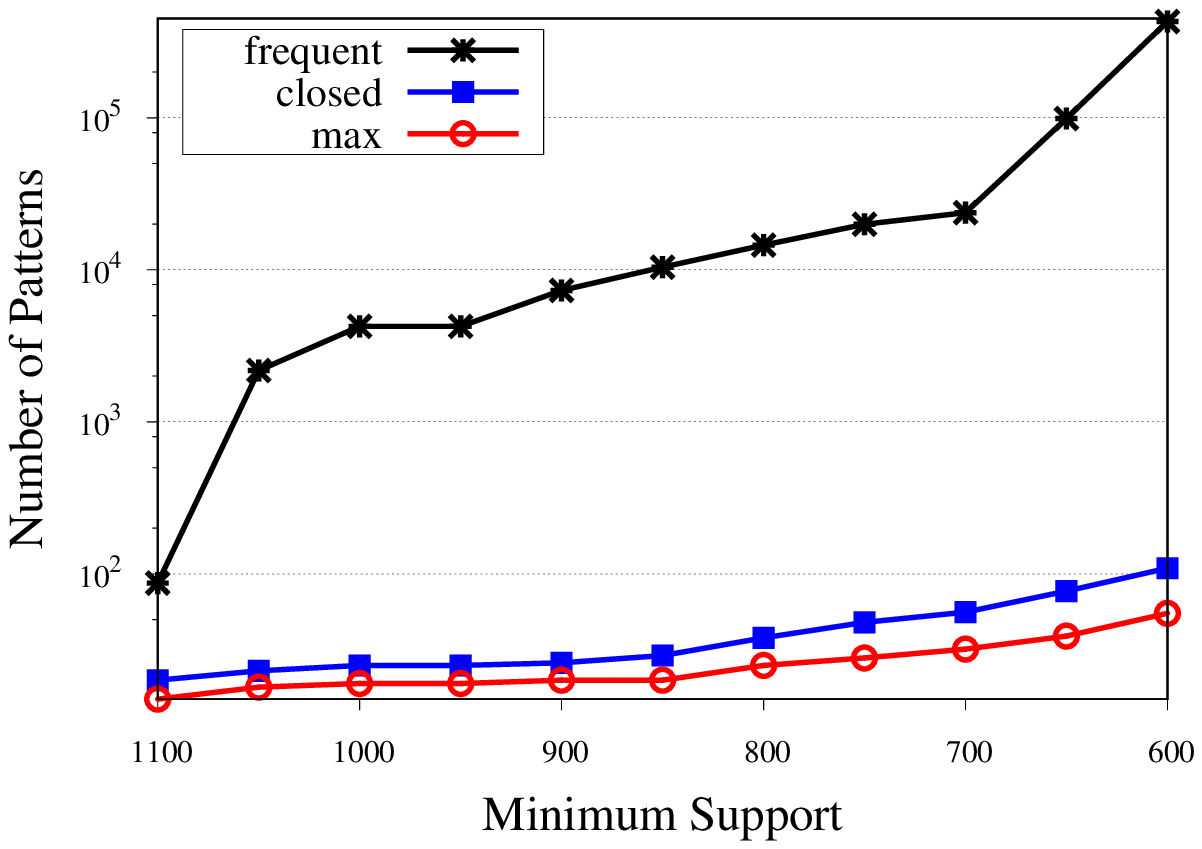}}}
        \vspace*{-3ex}
        \caption{Pattern distributions on XMark.}
      \label{fig:xmpatdist}
      \vspace*{-3ex}
\end{figure*}

\begin{figure*}[!t]
       \center
       \subfigure[Number of patterns vs. sizes]{ \scalebox{0.45}{ \label{fig:dppatlen} \epsfig{file=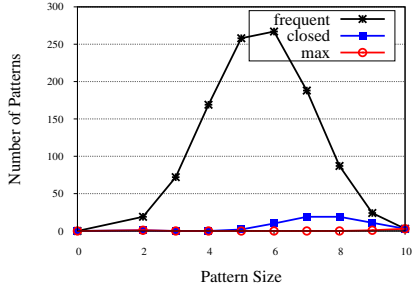} } }
       \subfigure[Number of patterns vs. minsup]{ \scalebox{0.45}{ \label{fig:dppatsup} \epsfig{file=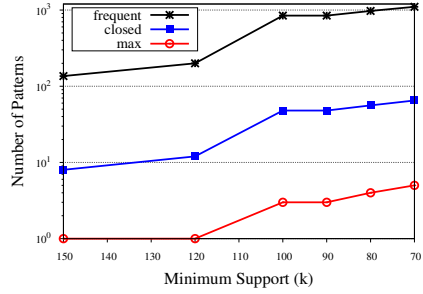}}}
        \vspace*{-3ex}
        \caption{Pattern distributions on DBLP.}
      \label{fig:dppatdist}
      \vspace*{-3ex}
\end{figure*}

\begin{figure*}[!t]
       \center
       \subfigure[Number of patterns vs. sizes]{ \scalebox{0.47}{ \label{fig:cspatlen} \epsfig{file=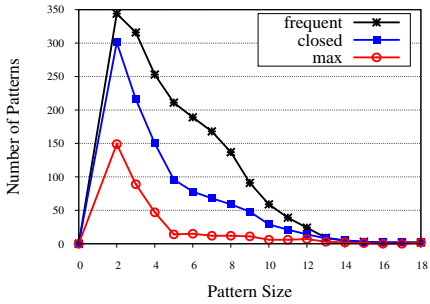} } }
        \hspace*{-.3cm}
       \subfigure[Number of patterns vs. minsup]{ \scalebox{0.45}{ \label{fig:cspatsup} \epsfig{file=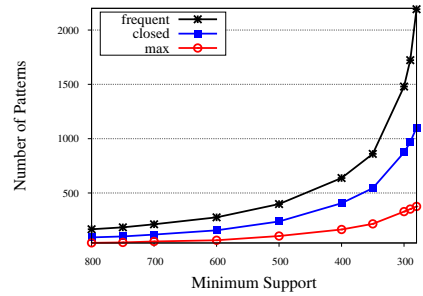}}}
        \vspace*{-3ex}
        \caption{Pattern distributions on CSlogs.}
      \label{fig:cspatdist}
\end{figure*}

\vspace*{.5ex}
\noindent\textbf{Pattern distribution.} To get a better understanding of  the characteristics of the datasests, we show the distribution of frequent, closed, and maximal embedded patterns on the three datasets in Figs. \ref{fig:xmpatdist}, \ref{fig:dppatdist}, and \ref{fig:cspatdist}. The distribution of frequent embedded patterns in relation to pattern size is shown in Figs. \ref{fig:xmpatdist}(a)-(b), \ref{fig:dppatdist}(a), and \ref{fig:cspatdist}(a) (the support threshold used is the lowest one considered on the three datasets). In the case of XMark, as the embedded patterns are subtantially more numerous than the closed and maximal patterns, we show the distributions of these patterns in two different figures (Figs. \ref{fig:xmpatdist}(a) and \ref{fig:xmpatdist}(b)). The number of frequent embedded patterns are shown in figures \ref{fig:xmpatdist}(c), \ref{fig:dppatdist}(b), and \ref{fig:cspatdist}(b). Different support thresholds on the three datasets are considered. {\em DBLP}  and {\em XMark} both exhibit an almost symmetric distribution. The mean values of the distributions though are different. The  distribution of {\em CSlogs} is right-skewed. 
Many  patterns in {\em CSlogs} are of small size. {\em XMark} has substantially larger frequent embedded patterns, and considerably more patterns than the other two datasets. The maximal patterns are contained in the closed patterns and both of them are contained in the set of frequent embedded patterns. We observe that, on  {\em XMark} and {\em DBLP}, the closed patterns and maximal patterns are substantially less numerous than the frequent embedded patterns; on {\em CSlogs}, the number differences among the three pattern types are moderate. As we show below, the distribution of patterns affects the effectiveness of the closed pattern pruning rules.

\subsection{Execution time and memory usage.}
\label{subsec:performance} 
\begin{figure*}[!t]
       \center
        \subfigure[Run time for closed \& max.  patterns]{ \scalebox{0.47}{ \label{fig:xmtimecl} \epsfig{file=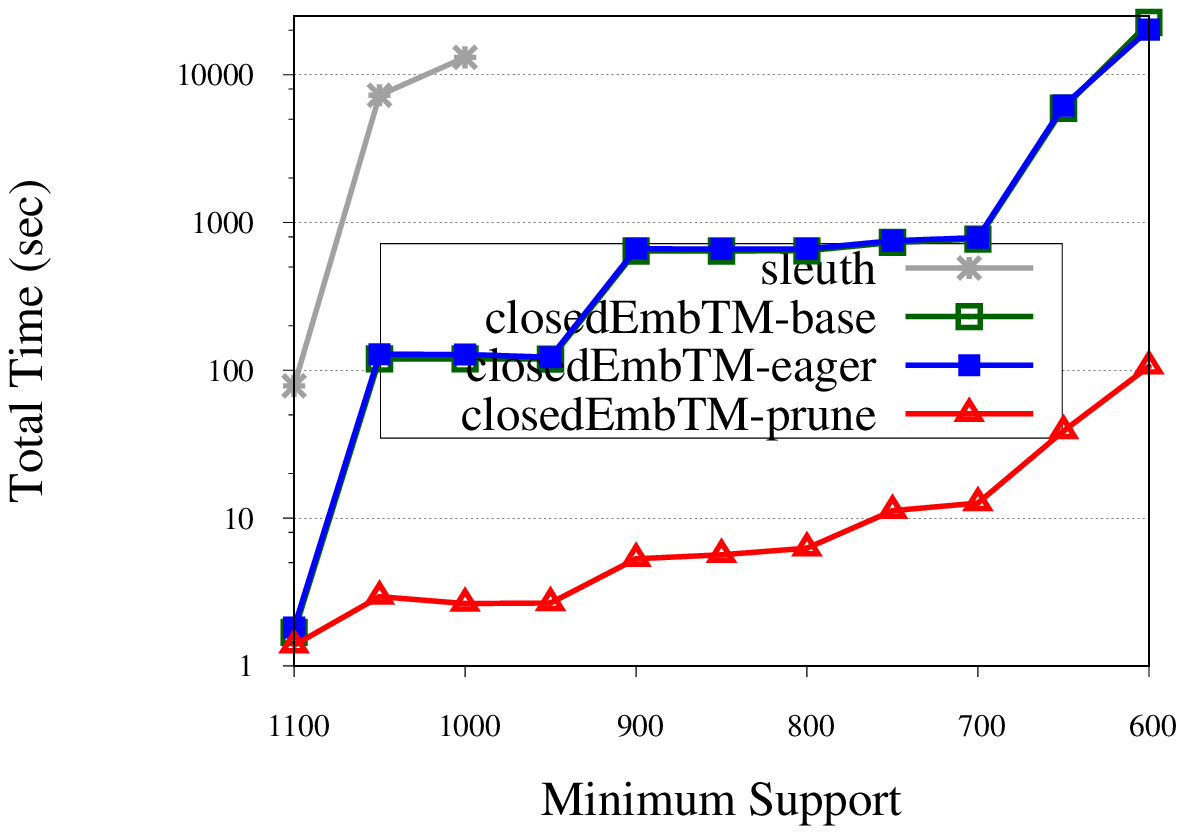} } }
        \hspace*{-.4cm}
        \subfigure[Memory usage for closed \& max. patterns]{ \scalebox{0.46}{ \label{fig:xmmemcl} \epsfig{file=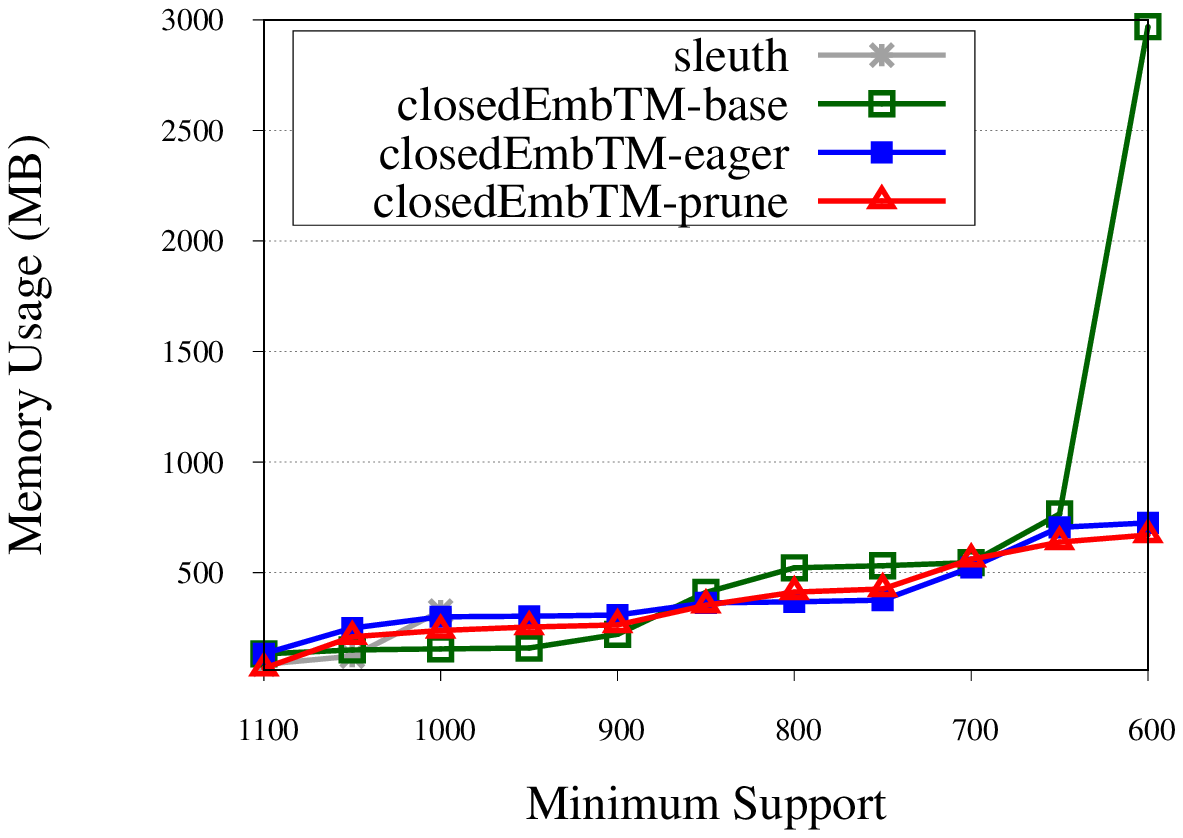}}}
         \vspace*{-2ex}
        \caption{Performance comparison on XMark for mining closed and maximal patterns.}
      \label{fig:xmperEmbCL}
\end{figure*}

\begin{figure*}[!t]
       \center
        \subfigure[Run time for closed \& max. patterns]{ \scalebox{0.47}{ \label{fig:dptimecl} \epsfig{file=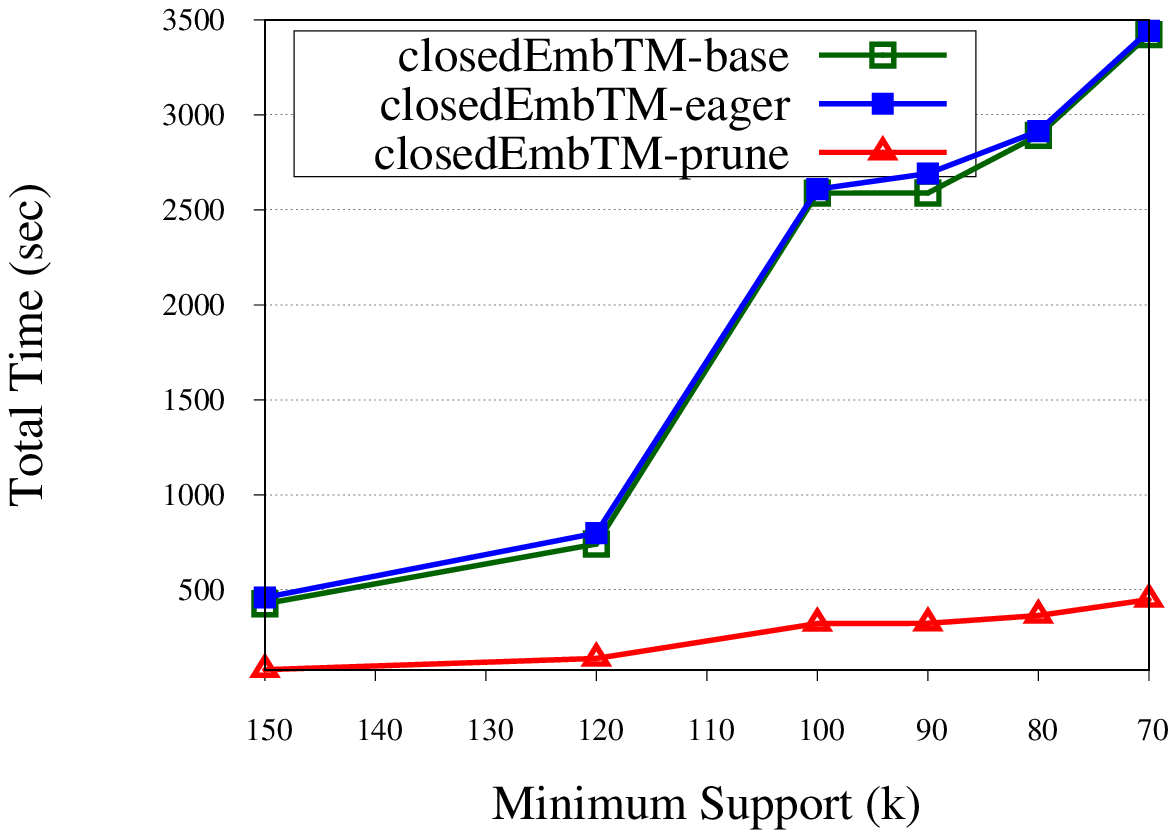} } }
        \hspace*{-.4cm}
        \subfigure[Memory usage for closed \& max. patterns]{ \scalebox{0.46}{ \label{fig:dpmemcl} \epsfig{file=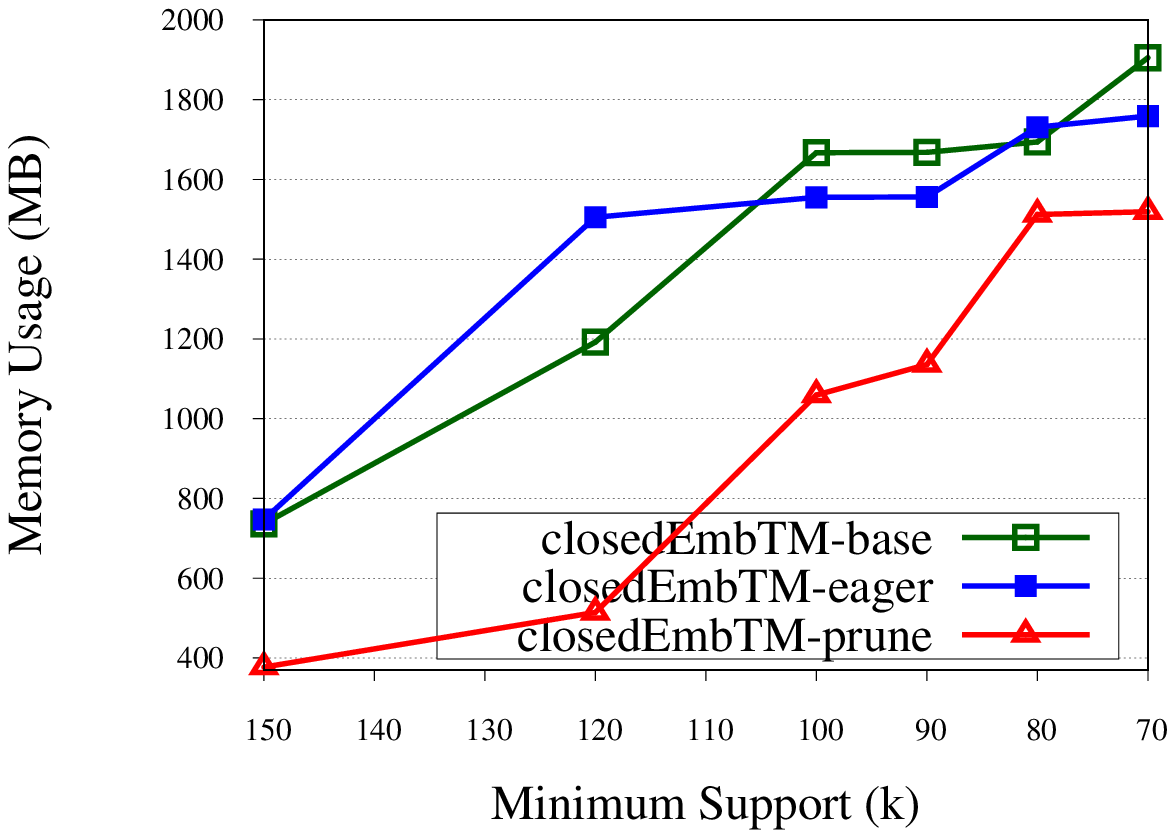}}}
         \vspace*{-2ex}
        \caption{Performance comparison on DBLP for mining closed and maximal patterns.}
      \label{fig:dpperEmbCL}
      \vspace*{-2ex}
\end{figure*}

\begin{figure*}[!t]
       \center
        \subfigure[Run time for clo. \& max. patterns]{ \scalebox{0.45}{ \label{fig:cstimecl} \epsfig{file=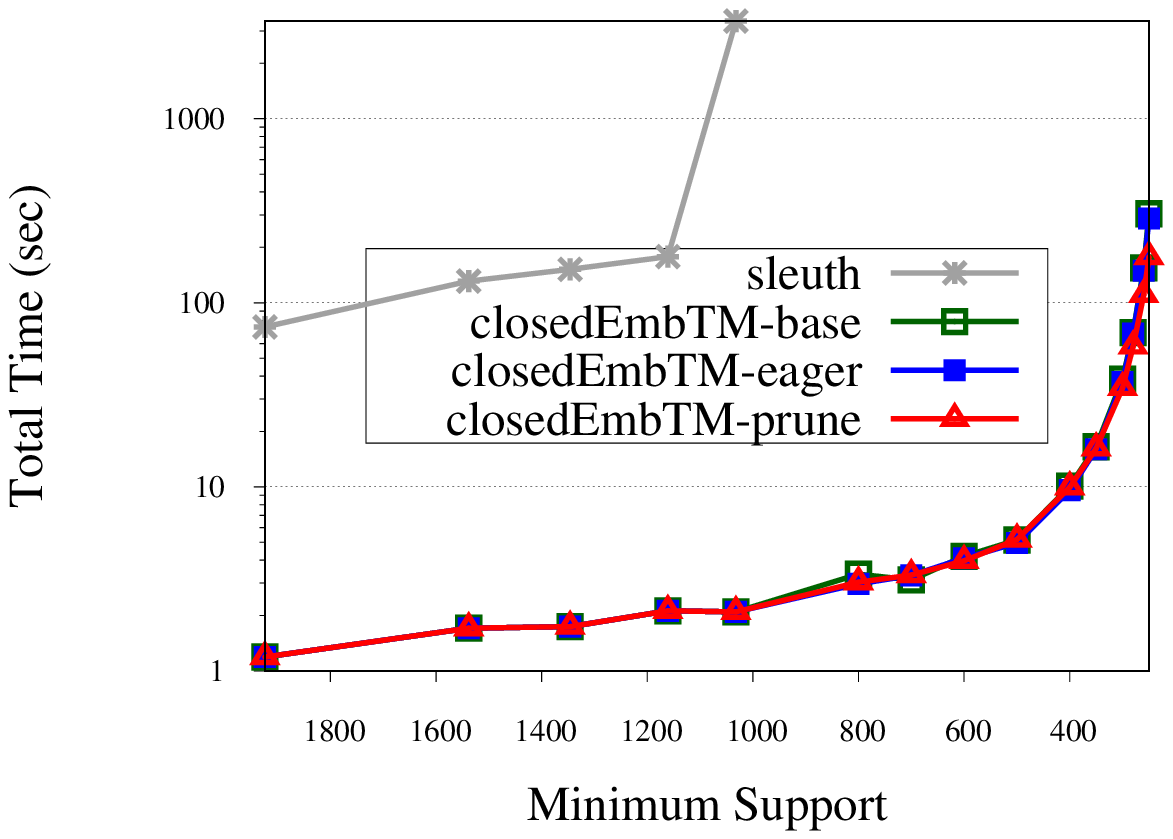} } }
        \subfigure[Memory usage for closed \& max. patterns]{ \scalebox{0.45}{ \label{fig:csmemcl} \epsfig{file=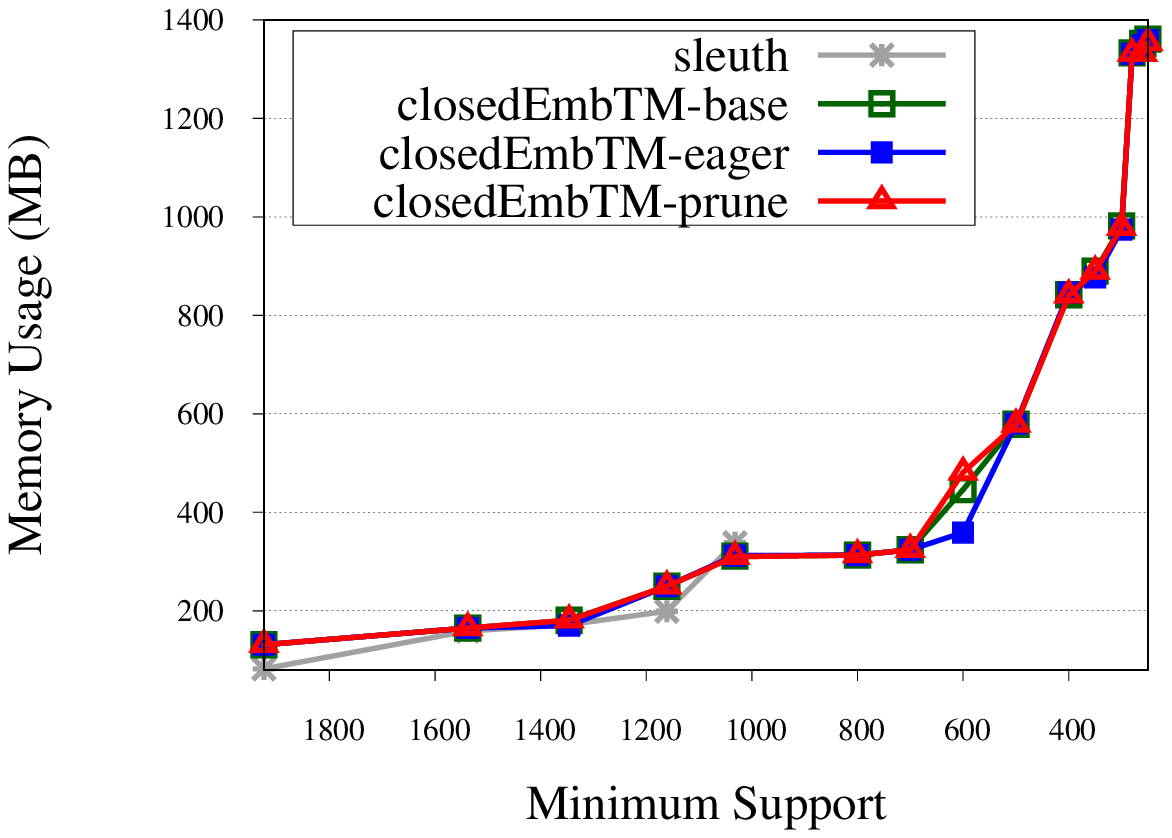}}}
         \vspace*{-2ex}
        \caption{Performance comparison on CSlogs for mining closed and maximal patterns.}
      \label{fig:csperEmbCL}
\end{figure*}

Figures \ref{fig:xmperEmbCL}, \ref{fig:dpperEmbCL} and \ref{fig:csperEmbCL} show the the total elapsed time and the memory footprint varying the support threshold of the four algorithms on XMark, DBLP and CSlogs, respectively. The Y-axis of Figs. \ref{fig:xmtimecl} and \ref{fig:cstimecl} is in a logarithmic scale. Table \ref{tab:patternstats} shows statistics for patterns generated and evaluated by {\em closedEmbTM-prune} and {\em closedEmbTM-eager} at the two smallest support levels tested on each dataset. The column {\em \#computed} records the number of generated patterns whose support has been computed. Columns {\em \#lclosed} and {\em \#lmax} record the number of locally closed and locally maximal embedded patterns, respectively. Note that {\em closed\-Emb\-TM-base} computes the same set of frequent and candidate patterns as {\em closedEmbTM-eager}, but it does not generate locally closed and maximal patterns. The following observations can be made.

\begin{table*}[!t]
\centering%
\begin{scriptsize}
\begin {tabular}[b]{|c|c|c|c|c|c|c|c|c|c|c|c|}
\hline
\multirow {2}*{dataset} & \multirow {2}*{minsup} &\multirow {2}*{\#closed} & \multirow {2}*{\#max.} & \multicolumn {4}{|c|}{algorithm {\em cEmbTM-eager}} & \multicolumn {4}{|c|}{algorithm {\em cEmbTM-prune}} \\
\cline {5-12}
&&&&\#computed &\#freq. &\#lclosed &\#lmax. &\#computed &\#freq. &\#lclosed &\#lmax. \\
\hline
\hline
 \multirow {2}*{XMark}& 650 & 59 & 34 & 857820	&74722	&21064	&20343	&9370	&720	&148 &123\\\cline {2-12}
 & 600	&83	&49	&3834456	&334220	&87146	&86146	&30283	&286	&2080	&240\\\cline {2-12}
 \hline
\multirow {2}*{DBLP}& 80k	&56	&4&	10836	&973	&668&384 &2480&186&60&30\\\cline {2-12}
 & 70k	&65	&5	&12282	&1103	&745	&433	&2782	&213	&69	&35\\\cline {2-12}
\hline
\multirow {2}*{CSlogs}& 260	&1479	&458	&463151	&3972	&2874	&1255	&458891	&2985	&2081	&915\\\cline {2-12}
 & 250	&1732	&502	&488113	&7188	&5020	&2336	&474364	&4448	&3008	&1306\\\cline {2-12}
\hline
\end {tabular}
\end{scriptsize}
\vspace*{-3ex}
\caption{Statistics for patterns computed by the algorithms in comparison.}
\label{tab:patternstats}
\end{table*}

First of all, {\em sleuth} runs orders of magnitude slower than the three algorithms\linebreak {\em closedEmbTM-base}, {\em closedEmbTM-eager} and  {\em closedEmbTM-prune}. On DBLP, {\em sleuth} cannot complete the computation within 12 hours not even at the largest support level tested. In addition, the rate at which the execution time of the three algorithms increases as the support level decreases is slower compared to {\em sleuth}. The explanation for this big performance difference is not the fact that {\em sleuth} computes embedded patterns which are more numerous than closed patterns. Instead, it is the fact that our algorithms calculate the pattern support more efficiently. {\em sleuth} is forced to record all  embedded occurrences of each considered pattern in the data, and join them. As a consequence, its performance suffers when mining dense data at low support levels. {\em sleuth} is also forced to keep in memory the occurrences of the considered patterns and this negatively affects memory consumption. Nevertheless, when the number of embedded occurrences at the corresponding support level is not large, {\em sleuth} does not consume too much memory and can complete the computation within a reasonable amount of time.

Second, {\em closedEmb\-TM-prune} outperforms {\em closedEmb\-TM-base} and {\em closedEmb\-TM-eager} on both XMark and DBLP by a great margin, with a speeding factor of up to 212 and 8 respectively. The large performance gap is due to the large number of candidate patterns pruned by {\em closedEmb\-TM-prune} on the two datasets. For instance, as shown in Table \ref{tab:patternstats}, on XMark at $minsup=600$,  the candidate patterns enumerated and the frequent patterns computed by {\em closedEmb\-TM-prune} are respectively 37 times and 49 times less than with {\em closedEmb\-TM-eager} (and {\em closedEmb\-TM-base}).  Also, comparing to {\em closedEmb\-TM-eager}, {\em closedEmb\-TM-prune} reduces the number of locally closed (resp. locally maximal) patterns that need to go through the closedness (resp. maximality) checking by a factor of 76 (resp. 87). On DBLP, the number of patterns checked and the number of frequent patterns computed by {\em closedEmb\-TM-prune} are also significantly smaller than with the other two methods.

The large gap in the number of generated patterns explains also the smaller memory usage by {\em closedEmbTM-prune} on both XMark and DBLP (Figs. \ref{fig:xmmemcl} and \ref{fig:dpmemcl}).  The advantage of memory consumption by {\em closedEmbTM-prune} is more prominent on DBLP. The reason is that DBLP contains a large number of patterns having repeated siblings. Checking the sibling constraint is more costly for patterns with repeated siblings. {\em closedEmbTM-prune} is able to skip computing the support of many of those patterns.  These results demonstrate the effectiveness of the pruning techniques used by {\em closedEmbTM-prune}, which allow skipping a large number of non-closed patterns from the pattern search space.

On CSlogs, {\em closedEmbTM-prune} runs slightly slower than {\em closedEmbTM-eager} until $minsup$ reaches 400. After that, {\em closedEmbTM-prune} catches up and runs faster as $minsup$ becomes smaller, with a speedup factor of 2 when $minsup$ is 200. A similar trend can be observed on memory usage: {\em closedEmbTM-prune} consumes a bit more memory than {\em closedEmbTM-eager} when $minsup$ exceeds 400; after that, its memory usage becomes similar to that of {\em closedEmbTM-eager}. This  can be explained by the fact that finding the child and cousin surrogates for the pruning rules incur an overhead. When the benefit from the reduction on the number of patterns considered due to pruning exceeds the pruning overhead, {\em closedEmbTM-prune} outperforms {\em closedEmbTM-eager}.

The time and memory performance of {\em closedEmbTM-eager} and {\em closedEmbTM-base} are in general very close. However, {\em closedEmbTM-eager} has a noticeable performance advantage over {\em closedEmbTM-base} once the support level becomes very small. Note that the frequent patterns generated by the two algorithms are exactly the same. By maintaining a set of locally-closed patterns, {\em closedEmbTM-eager} succesfully reduces the total number of pair-wise pattern containment checking to identify closed patterns.  With {\em closedEmbTM-base}, the entire set of frequent patterns, whose size can be exponential in the size of the locally-closed pattern set, is checked for closedness. In general, when the frequent pattern set examined is not very large, the total time spent on pattern closedness checking is very small. For instance, on XMark, when the frequent patterns are less than 100k, the closedness checking time does not even represent 2.6\% of the total execution time of {\em closedEmbTM-base}. However, this percentage jumps to 31.6\% when $minsup$ is 600 and the number of frequent patterns is around 430k.  Also, at this point, the memory consumption of {\em closedEmbTM-base} exceeds that of {\em closedEmbTM-eager} by a factor of over 7.

\begin{figure*}[!t]
       \center
       \subfigure[Run time]{ \scalebox{0.45}{ \label{fig:xmtimescale} \epsfig{file=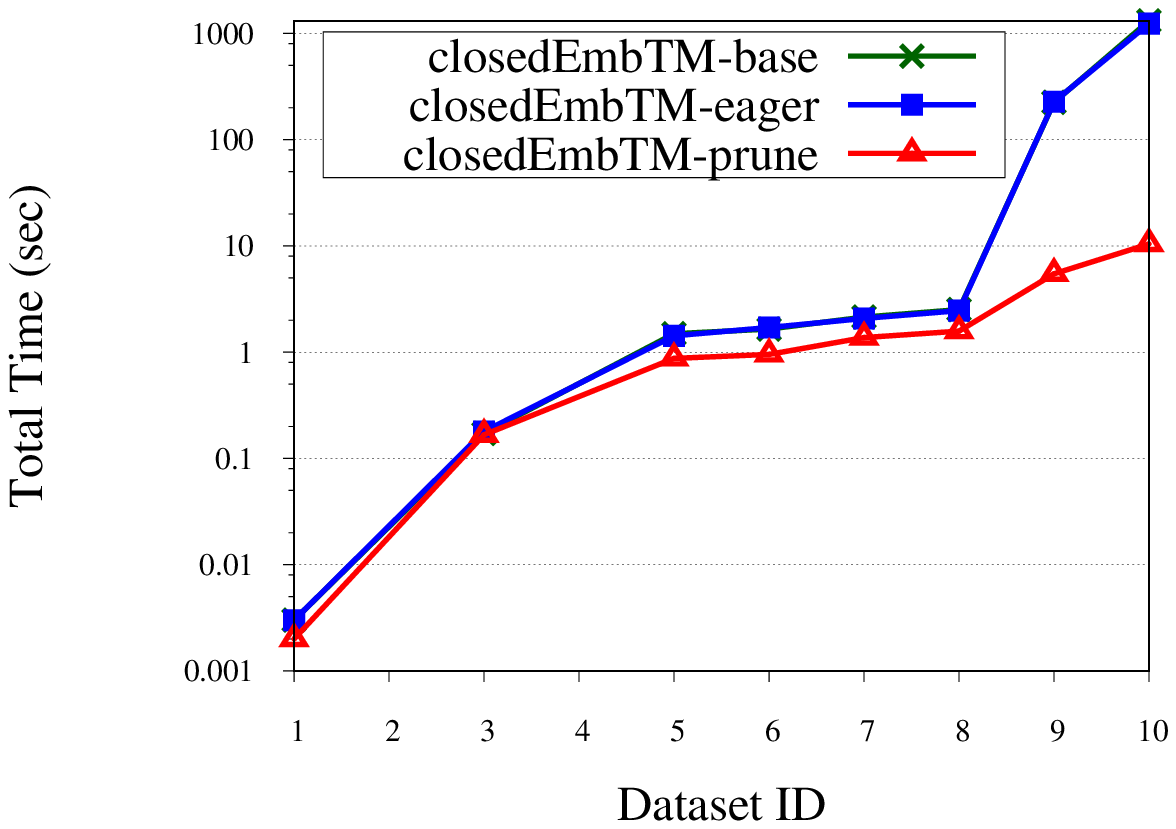} } }
        \subfigure[Memory usage]{ \scalebox{0.45}{ \label{fig:xmmemscale} \epsfig{file=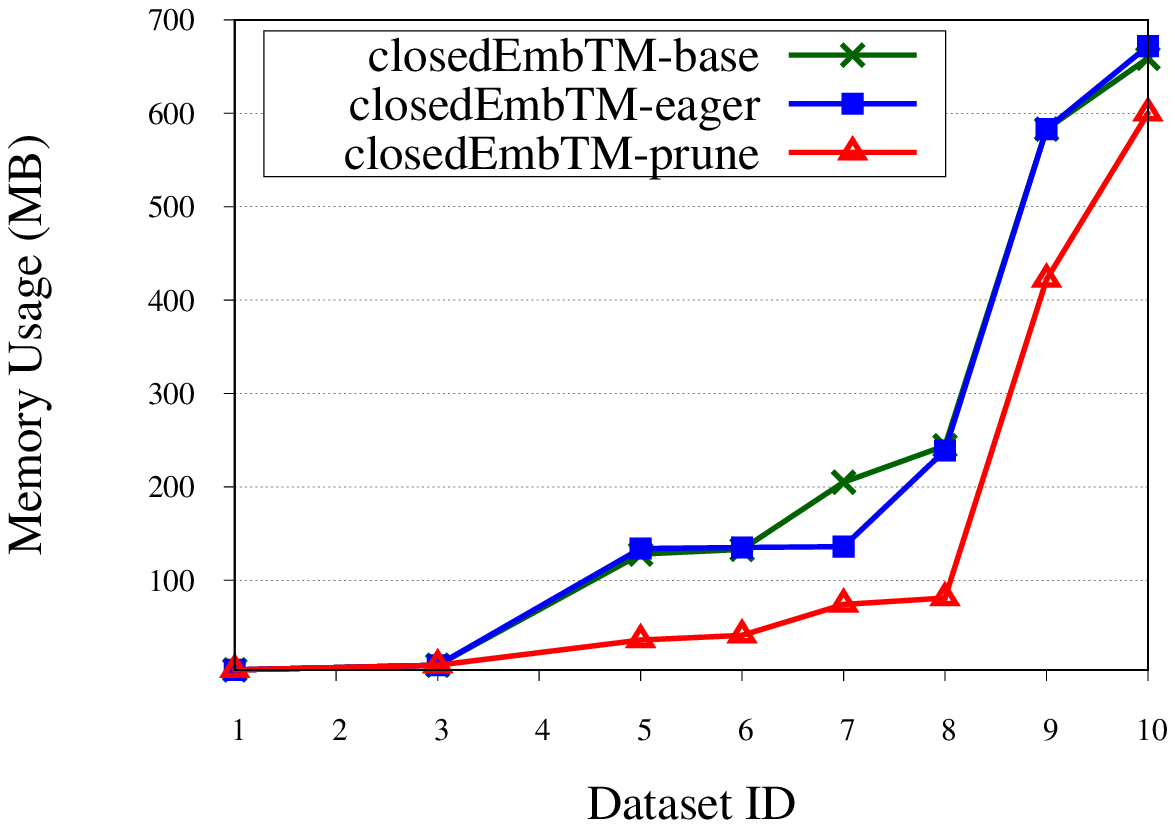}}}
        \vspace*{-3ex}
        \caption{Scalability comparison for mining closed and maximal patterns on XMark with increasing size ($minsup$ = 1800).}
      \label{fig:xmscale}
      \vspace*{-3ex}
\end{figure*}

\begin{figure*}[!t]
       \center
       \subfigure[Run time]{ \scalebox{0.45}{ \label{fig:cstimescale} \epsfig{file=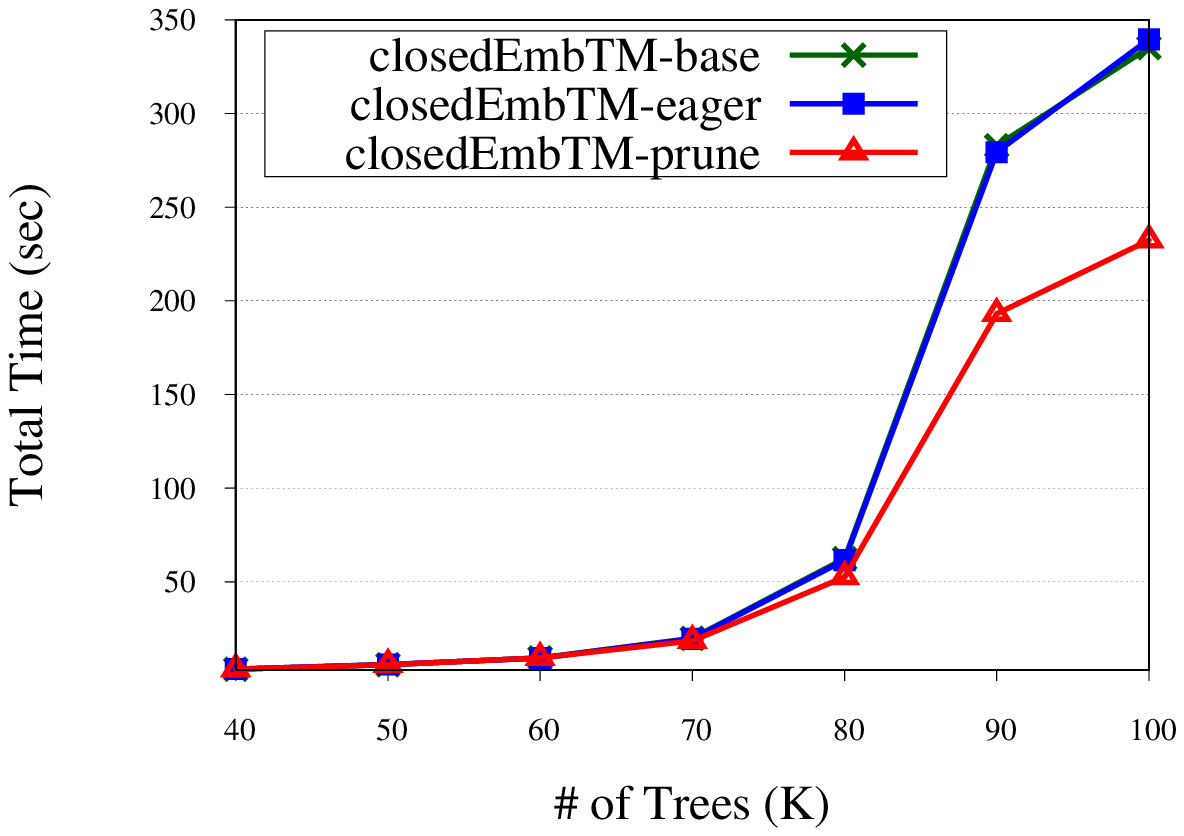} } }
        \subfigure[Memory usage]{ \scalebox{0.45}{ \label{fig:csmemscale} \epsfig{file=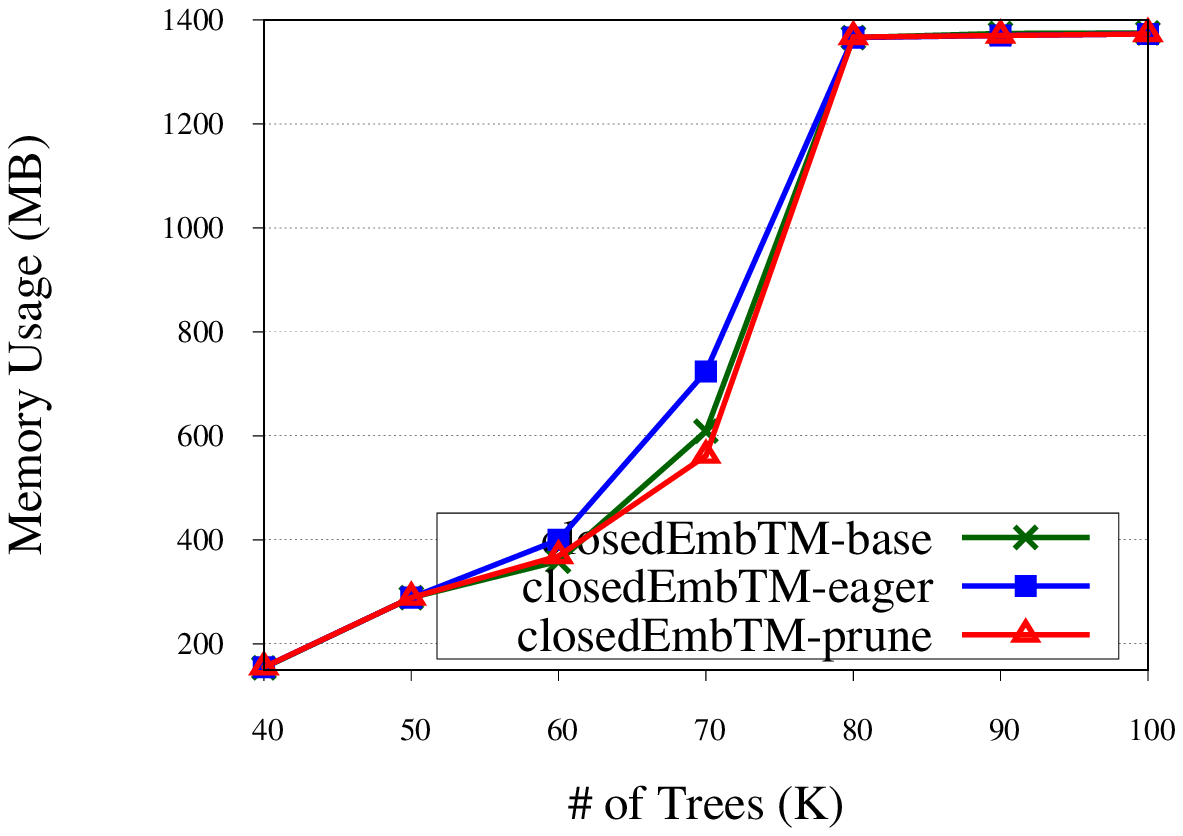}}}
        \vspace*{-3ex}
        \caption{Scalability comparison for mining closed and maximal patterns on CSlogs with increasing size ($minsup$ = 400).}
      \label{fig:csscale}
      \vspace*{-2ex}
\end{figure*}

\subsection{Scalability Comparison}
\label{subsec:scalability}
We studied the scalability of {\em closedEmbTM-prune}, {\em closedEmbTM-eager} and \linebreak{\em closedEmbTM-base} as the size of the input data on XMark (which is a single large data tree) and CSlogs (which is a collection of small trees) increases. For the experiments with XMark, 10 XMark trees were generated. The {\em scaling factor} parameter of XMark was set to 0.01, 0.02, \ldots, 0.10 and the $minsup$ was fixed at 1800. For the experiments with CSlogs, seven datasets of different sizes (ranging from 40k to 100k trees) were generated. The data trees were randomly selected from the original CSlogs dataset. The $minsup$ was fixed at 400.

The results on XMark (Fig. \ref{fig:xmscale}) show that {\em closedEmbTM-base} and {\em closedEmbTM-eager} have similar time performance and memory usage, and\linebreak {\em closedEmbTM-prune}  has the best time performance. The execution time of \linebreak{\em closedEmbTM-prune} progresses smoothly as the input data size increases, whereas the execution time of the other two algorithms increases sharply. Overall, \linebreak{\em closedEmbTM-prune} has a speedup factor up to  125 over the other two. Its memory consumption is also smaller than the other two in all the testing cases.

The results on CSlogs (Fig. \ref{fig:csscale}) show that {\em closedEmbTM-base} and {\em closedEmbTM-eager} have similar time performances, while {\em closedEmbTM-prune} runs noticeably faster than the other two when the dataset size reaches and goes above 70k. The three methods have overall similar memory performances.

%% file: texs/related.tex
\section{Related Work}
\label{sec:related}

We review in this section related literature and comment on how it relates to our work. Because of its importance, the problem of extracting tree patterns from collections of small trees has attracted a lot of attention in recent years. Different algorithms have been proposed  \cite{AsaiAKASA02,TermierRS02,AsaiAUN03,XiaoYLD03,ChiYM04,NijssenK04,FengHL05,ChiXYM05,Zaki05,Zaki05tkde,GoethalsHB05,TatikondaPK06,TermierRSOWM08,TanHDCF08,DriesN12,KibriyaR13,WuT15,WuTSBDR18}, but only a handful of them address the problem of mining unordered embedded patterns \cite{TermierRS02,FengHL05,Zaki05,WuT15,WuTSBDR18}. None of them introduces closed or maximal unordered embedded tree patterns or discusses pattern summarization.

There are some algorithms \cite{XiaoYLD03,ChiXYM05,OzakiO06,TermierRSOWM08} which deal with extracting closed or maximal {\em induced} patterns. Recall that induced patterns are based on isomorphisms which, contrary to embeddings,  cannot map pattern edges to paths in the data. Further, these algorithms are restricted to extracting patterns from collections of small trees as opposed to extracting patterns from a single large tree. We briefly review these approaches below.

$CM\-Tree\-Miner$ \cite{ChiXYM05} mines both closed and maximal frequent patterns from a set of small trees. This approach exploits a concept called {\em blanket} of a pattern which is the set of all its frequent immediate super patterns. The occurrences of a pattern under consideration are compared with those of its blanket pattern to determine whether it is closed or not. In order to reduce the search space, $CM\-Tree\-Miner$ employs heuristic and pruning techniques.  Unlike previous algorithms, it precisely extracts maximal and closed patterns without previously producing all frequent patterns. Nevertheless, it is restricted on mining only induced patterns. 
Two mining algorithms for closed induced ordered patterns are described in \cite{OzakiO06}. These algorithms combine the pruning techniques of $CM\-Tree\-Miner$ with two different pattern enumeration strategies.

$PathJoin$ \cite{XiaoYLD03} operates on a collection of small trees and extracts maximal unordered induced patterns. $PathJoin$ accepts only datasets where sibling nodes in the data tree cannot be labeled by the same label. It proceeds by initially discovering all the maximal frequent paths. Candidate frequent tree patterns are then found by joining the maximal frequent paths. These candidate patterns are then used to select maximal frequent patterns through a post-processing filtering phase.  This method has a considerable performance penalty, in particular, when there are many false positive patterns.

$Dryade\-Parent$ \cite{TermierRSOWM08} also operates on collections of small trees and extracts closed induced patterns. It proceeds by first computing all closed frequent patterns of depth one which are called tiles. It then employs a hooking method which uses these tiles to construct closed frequent patterns. The authors observe that one parameter that dramatically affects the performance of previous algorithms is the outdegree of the tree patterns. Therefore, similar to $PathJoin$, it restricts the input data trees to those that do not have two sibling nodes with the same label. Even though this assumption certainly makes the problem easier, it also narrows the usability of the approach in real cases.

As we have explained in the introduction, the approaches above cannot be leveraged for mining closed and maximal {\em embedded} patterns from a {\em large tree}.

There has been only very limited contributions on mining tree patterns from a single tree or graph. They are restricted to \cite{GoethalsHB05,DriesN12,KibriyaR13} which focus on extracting from a graph patterns with child edges, and to \cite{WuT15,WuTSBDR18} which leverage homomorphisms to extract embedded patterns from a tree. We are not aware of any paper on efficiently summarizing embedded tree patterns from a single large data tree.

%% file: texs/conclusion.tex
\section{Conclusion}
\label{sec:conclusion}

Data from many modern applications are represented, exchanged and exported in tree-structured form. Large tree datasets are continuously produced by those applications. To produce knowledge from this type of data, we have studied the problem of summarizing frequent embedded tree patterns from large tree datasets. To the best of our knowledge, this problem has not been investigated by existing studies.

We have introduced the concepts of closed and maximal embedded unordered tree patterns and studied their properties in the setting of a single large data tree. We have designed a local closedness checking technique to eagerly eliminate non-closed patterns. We have further designed pattern search space pruning rules to proactively detect and prune patterns that do not correspond to closed ones. These techniques have been integrated into embedded frequent pattern mining algorithms in order to mine all the maximal and closed embedded frequent patterns from large  tree data.  The experimental results on synthetic and real datasets demonstrate that, on dense datasets, our approach generates a complete closed and maximal pattern set which is substantially smaller than that generated by the embedded pattern miner, but also runs much faster with negligible overhead on pattern pruning.

Our future work focuses on incorporating user-specified constraints to the proposed approach in order to further reduce the size of the result set and to enable constraint-based mining of compact sets of frequent embedded tree patterns.

%% file: texs/appendix.tex
\newpage\section*{APPENDIX}  

\comments{
\begin{lemma}
\label{lem:prefixOccu}
{\em Given two patterns $P$ and $P'$, if $P \sqsubseteq_e P'$, then $\Pi_P(P')\sqsubseteq OC^e(P).$}
\end{lemma}

\vspace{1.5ex}\noindent\textbf{Proof.} Since $P$ is an embedded subpattern of $P',$ by the definitions of embedded subpattern and pattern projection, for each $e\in \Pi_P(P'),$ we have $e\in OC^e(P).$ \hfill$\Box$
}

\vspace{1.5ex}\noindent\textbf{Proposition 3} {\em Consider two distinct patterns $P_{x}^i$ and $P_{y}^j, \; i \geq j,$ in the class $[P]$, such that $P_{x}^i < P_{y}^j$. If $P_{y}^j \equiv P_x^i \otimes_s P_y^j$, and for any $P_{z}^k\in[P]$ such that $P_{y}^j < P_{z}^k$, we have $i>k$, then  $P_{x}^i$ is a cousin surrogate of $P_{y}^j$.}

\vspace{1.5ex}\noindent\textbf{Proof.} Clearly condition (a) of Definition \ref{def:cnsurrogate} is satisfied since $P_{y}^j \equiv P_x^i \otimes_s P_y^j$. This occurrence equivalence implies that for every embedding $e$  of $P_{y}^j$ to a data tree $T$, there is also an embedding of  $P_x^i \otimes_s P_y^j$ to $T$ which is an extension of $e$ such that the parent of the image of node $y$ is an ancestor-or-self of the image of node $x$ in $T$ and the images of node $x$ and node $y$ are not on the same path in $T$.

To show that condition (b) of Definition \ref{def:cnsurrogate} is also satisfied, we first consider $P_y^j \otimes_c P_z^k$ (which means that $j=k$). Every embedding $e'$ of  $P_y^j \otimes_c P_z^k$ to $T$ maps node $z$ to a descendant of the image of node $y$ under $e'$ in $T$. For every embedding $e'$ of $P_y^j \otimes_c P_z^k$, a restriction $e''$ of $e'$ obtained by excluding the mapping of node $z$ is an embedding of $P_{y}^j$. Therefore, there is an extension of $e''$ which is also an embedding of  $P_x^i \otimes_s P_y^j$ to $T$ such that the images of node $x$ and node $y$ are not on the same path in $T$ (and therefore, the images of node $x$ and node $z$ are not on the same path in $T$). This extension is also an embedding of $(P_x^i \otimes_s P_y^j) \otimes_c (P_x^i \otimes_s P_z^k)$. That is, $P_y^j \otimes_c P_z^k \equiv (P_x^i \otimes_s P_y^j) \otimes_c (P_x^i \otimes_s P_z^k)$.

Let's now consider $P_y^j \otimes_s P_z^k$. Every embedding $e'$ of  $P_y^j \otimes_s P_z^k$ to $T$ maps the parent of node $z$ to an ancestor-or-self of the image of the parent of node $y$ under $e'$ in $T$. For every embedding $e'$ of $P_y^j \otimes_s P_z^k$, a restriction $e''$ of $e'$ obtained by excluding the mapping of node $z$ is an embedding of $P_{y}^j$. Therefore, there is an extension of $e''$ which is also an embedding of  $P_x^i \otimes_s P_y^j$ to $T$ such that the images of node $x$ and node $y$ are not on the same path in $T$. Since $i > k$, the image of node $x$ under $e''$ is not in the same path as the image of node $z$ under $e'$ in $T$. Consequently, this extension is also an embedding of $(P_x^i \otimes_s P_y^j) \otimes_s (P_x^i \otimes_s P_z^k)$. That is, $P_y^j \otimes_s P_z^k \equiv (P_x^i \otimes_s P_y^j) \otimes_s (P_x^i \otimes_s P_z^k)$.


\vspace{1.5ex}\noindent\textbf{Lemma 1} {\em If a pattern has a child or a cousin surrogate, then neither itself nor anyone of its (child or cousin) expansion outcomes can be closed.}

\vspace{1.5ex}\noindent\textbf{Proof.} If a pattern $P_{y}^j$ has a child or a cousin surrogate $P_{x}^i$, then by Condition (a) in Definition \ref{def:csurrogate} or \ref{def:cnsurrogate}, respectively, the pattern is not closed. Also, if a pattern $P_{y}^j$ has a child or a cousin surrogate $P_{x}^i$, then by Condition (b) in Definition \ref{def:csurrogate} or \ref{def:cnsurrogate}, respectively, for any $P_{z}^k\in[P]$ such that $P_{y}^j\leq P_{z}^k$, all the possible expansion outcomes on $P_{y}^j$ by $P_{z}^k$ are not closed.

To show that all the possible expansions on $P_{y}^j$ by $P_{z}^k$ are not closed also for any $P_{z}^k\in[P]$ such that $P_{y}^j > P_{z}^k$, we first observe that the cousin expansion on $P_{y}^j$ by $P_{z}^k$ is either not possible (when $k>j$), or the outcome is not canonical (when, $k=j$). The child expansion of $P_{y}^j$ by $P_{z}^k$  (which means that $k=j$) is split in two cases: 

The first case assumes that $P_x^i$ is a cousin surrogate of $P_y^j.$  Then, the child expansion of $P_{y}^j$ by $P_{z}^k$ can be shown to be non-closed by an argument similar to the one presented in the proof of Proposition \ref{prop:cnsurrogate}. 

The second case assumes that $P_x^i$ is a child surrogate of $P_y^j$ (which means that $i=j$). By condition (a) of Definition \ref{def:csurrogate} for every embedding $e$  of $P_{y}^j$ to a data tree $T$, there is also an embedding of  $P_x^i \otimes_s P_y^j$ to $T$ which is an extension of $e$ such that the parent of the image of node $y$ is a descendant of the image of node $x$ in $T$. Every embedding $e'$ of  $P_y^j \otimes_c P_z^k$ to $T$ (remember that we consider a child expansion of $P_{y}^j$ by $P_{z}^k$) maps node $z$ to a descendant of the image of node $y$ under $e'$ in $T$. But then, there is an extension $e''$ of $e'$ which maps node $x$ to an ancestor of node $y$ in $T$. Extension $e''$ is an embedding of $(P_x^i \otimes_c P_y^j) \otimes_c (P_x^i \otimes_c P_z^k)$, i.e.,  $P_y^j \otimes_c P_z^k \equiv (P_x^i \otimes_c P_y^j) \otimes_c (P_x^i \otimes_c P_z^k)$. This implies that $P_y^j \otimes_c P_z^k$ is non-closed. Therefore, all the possible expansions on $P_{y}^j$ by $P_{z}^k$ are not closed also for any $P_{z}^k\in[P]$ such that $P_{y}^j > P_{z}^k$.


\vspace{1.5ex}\noindent\textbf{Proposition 4} {\em If a pattern has a child or a cousin surrogate, then neither itself nor anyone of its descendants in the pattern search tree can be closed (or maximal).}

\vspace{1.5ex}\noindent\textbf{Proof.} Let $P_{x}^i$ be a (child or cousin) surrogate of $P_{y}^j.$ In order to prove Proposition~4, it suffices to prove that any descendant pattern $Q$ of $P_{y}^j$ in the pattern search tree has a superpattern $Q',$  obtained by augmenting $Q$ with a node $x$, such that $Q \equiv Q'$. If $P_{x}^i$ is a child surrogate of $P_{y}^j$, node $x$ is added in $Q$ as the parent of node $y$. If $P_{x}^i$ is a cousin surrogate of $P_{y}^j$, node $x$ is added in $Q$ as the direct left sibling of node $y$ (when $i=j$), or as the leaf node on the path from the  left sibling of node $y$ (when $i>j$).  We will prove this claim by induction on the number of nodes $m$ added to $P_{y}^j$ to construct $Q.$

\vspace*{1ex}\noindent\textbf{Base Case:} $m=1.$ The base case corresponds to Lemma \ref{lem:rule1} which has been proven above.

\vspace*{1ex}\noindent\textbf{Inductive Case:} Assuming that the proposition holds for $m\leq n-1$ ($n\geq 2$), we prove next that it also holds for $m=n.$ Let node $z$ be the $m$-th node added to $P_{y}^j$ resulting in pattern $Q$. Let also $S$ be the pattern before the addition of $z.$

By the inductive assumption, we know that there is some pattern $S'$, obtained by augmenting $S$ with a node $x$, such that $S \equiv S'$.  We construct a pattern $Q'$ by attaching $z$ to $S'$ in the same position as $z$ was added in $Q.$ We show below that $Q \equiv Q'$.




Consider an embedding $e$ of $Q$ extended to a mapping $e'$ which maps node $x$ to $T$ so that its restriction obtained by removing node $z$ is an embedding for $S'$. This is possible by the induction hypothesis. We show next that $e'$ is an embedding also  for $Q'$. In order to do so, we need to prove that if nodes $x$ and $z$ are not on the same path in $Q'$, their images under $e$ in $T$ are not on the same path either.  If nodes $x$ and $z$ are not on the same path in $Q'$, they can have three relative positions in $Q'$: (1)~$z$ is a descendant of the rightmost sibling of $x$; (2)~$z$ is a descendant of an ancestor of the parent of $x$; and (3) $z$ is the rightmost sibling of $x$. Clearly, in the first two cases, the images of $x$ and $z$ under $e$ cannot be on the same path. Condition (b) in Definition \ref{def:csurrogate} and \ref{def:cnsurrogate} guarantees that they cannot be on the same path in the third case either. Hence, $e'$ is an embedding of $Q'$. Therefore, $Q \equiv Q'$.

\vspace{1.5ex}\noindent\textbf{Proposition 5} {\em If $P_{x}^i$ is a child surrogate of $P_{y}^i$ then for any $P_{z}^i\in[P]$ such that $P_{y}^i<P_{z}^i$ and $\Pi_{P}(P_{y}^i) = \Pi_{P}(P_z^i)$, $\Pi_{P_{x}^i}(OC(P_x^i \otimes_c P_y^i))\subseteq \Pi_{P_{x}^i}(OC(P_{x}^i \otimes_c P_{z}^i))$ or $\Pi_{P_x^i}(OC(P_x^i \otimes_c P_y^i))\subseteq \Pi_{P_x^i}(OC(P_x^i \otimes_s P_z^i))$.}

\vspace{1.5ex}\noindent\textbf{Proof.} We prove the proposition by contradiction. Suppose that $P_{x}^i$ is a child surrogate of $P_{y}^i$ but there exists $P_{z}^i\in[P]$ such that $P_{y}^i<P_{z}^i$, $\Pi_{P_{x}^i}OC((P_x^i \otimes_c P_y^i))\not\subseteq \Pi_{P_{x}^i}(OC(P_{x}^i \otimes_c P_{z}^i))$ and $\Pi_{P_x^i}(OC(P_x^i \otimes_c P_y^i))\not\subseteq \Pi_{P_x^i}(OC(P_x^i \otimes_s P_z^i))$. That is, there exists an occurrence of $P_{x}^i$ which is part of an occurrence of $P_x^i \otimes_c P_y^i$ but is not part of any occurrence of $P_{x}^i \otimes_c P_{z}^i$, and there exists a (not necessarily distinct)  occurrence of $P_{x}^i$ which is part of an occurrence of $P_x^i \otimes_c P_y^i$, but it is not part of an occurrence of $P_{x}^i \otimes_s P_{z}^i$. But then, $(P_y^i \otimes_s P_z^i) \not\equiv (P_x^i \otimes_c P_y^i) \otimes_s (P_x^i \otimes_c P_z^i)$ and $(P_y^i \otimes_s P_z^i) \not\equiv (P_x^i \otimes_c P_y^i) \otimes_s (P_x^i \otimes_s P_z^i)$. That is, condition (b) of Definition \ref{def:csurrogate} is violated. This means that $P_{x}^i$ is not a child surrogate of $P_{y}^i$, a contradiction.\hfill$\Box$

\vspace{1.5ex}\noindent\textbf{Proposition 6} {\em  If $P_{x}^i$ is a cousin surrogate of $P_{y}^i$, then for any $P_{z}^{i}\in[P]$ such that $P_{y}^{i} < P_{z}^{i}$, $\Pi_{P_{x}^{i}}(OC(P_x^i \otimes_s P_y^i))$ is disjoint from both $\Pi_{P_{x}^{i}}(OC(P_{x}^{i} \otimes_c P_{z}^{i}))$ and $\Pi_{P_{x}^{i}}(OC(P_{z}^{i} \otimes_c P_{x}^{i})).$}

\vspace{1.5ex}\noindent\textbf{Proof.} We prove it by contradiction. Suppose that $P_{x}^i$ is a cousin surrogate of $P_{y}^i$ and there exists $P_{z}^{i}\in[P]$, $P_{y}^{i} \leq P_{z}^{i}$, such that $\Pi_{P_{x}^{i}}(OC(P_x^i \otimes_s P_y^i))$  intersects with $\Pi_{P_{x}^{i}}(OC(P_{x}^{i} \otimes_c P_{z}^{i}))$ or $\Pi_{P_{x}^{i}}(OC(P_{z}^{i} \otimes_c P_{x}^{i})).$ Then, there exists an image of $z$ under an embedding of $P_{x}^{i} \otimes_c P_{z}^{i}$ or of $P_{z}^{i} \otimes_c P_{x}^{i}$ to the data graph  which is on the same path with the image of $x$ under an embedding of $P_x^i \otimes_s P_y^i.$  That is, condition (b) of Definition \ref{def:cnsurrogate} is violated. This means that $P_{x}^i$ is not a cousin surrogate of $P_{y}^i$, a contradiction. \hfill$\Box$